\documentclass[aps,prx,singlecolumn,groupedaddress,showpacs,superscriptaddress,amssymb,amsmath]{revtex4-2}
\usepackage{graphicx}
\usepackage{cancel}
\usepackage{tabularx}
\usepackage{color}
\usepackage{amsmath}
\usepackage{amsfonts}
\usepackage{amssymb}
\usepackage{comment}
\usepackage{verbatim}
\usepackage{amsthm}

\newcommand{\be}{\begin{equation}}
\newcommand{\ee}{\end{equation}}
\newcommand{\bea}{\begin{eqnarray}}
\newcommand{\eea}{\end{eqnarray}}
\usepackage{dcolumn}
\usepackage{hyperref}
\usepackage{bm}
\usepackage{epsf}
\usepackage{subfig}
\usepackage{epstopdf}
\usepackage{amsfonts}

\newcommand{\unit}{1\!\!1}
\newcommand{\DDD}{\color{magenta}}
\newcommand{\AAA}{\textcolor{blue}}
\usepackage{amsthm}
\theoremstyle{definition}

\begin{document}

\title{Collisional model with dissipative and dephasing baths: Nonadditive effects 
at strong coupling}


\author{Alessandro Prositto}

\email{alessandro.prositto@mail.utoronto.ca}

\affiliation{Department of Physics and Centre for Quantum Information and Quantum Control, University of Toronto, 60 Saint George St., Toronto, Ontario, M5S 1A7, Canada}

\author{Carlos Ramon-Escandell}
\affiliation{Department of Physics and Centre for Quantum Information and Quantum Control, University of Toronto, 60 Saint George St., Toronto, Ontario, M5S 1A7, Canada}

\author{Dvira Segal}
\email{dvira.segal@utoronto.ca}
\affiliation{Department of Chemistry, University of Toronto, 80 Saint George St., Toronto, Ontario, M5S 3H6, Canada}
\affiliation{Department of Physics and Centre for Quantum Information and Quantum Control, University of Toronto, 60 Saint George St., Toronto, Ontario, M5S 1A7, Canada}

\date{\today}
\begin{abstract}
The repeated interaction model provides a framework for emulating and analyzing the dynamics of open quantum systems. We explore here the dynamics generated by this protocol in a system that is simultaneously coupled to two baths through noncommuting system operators. One bath is made to couple to nondiagonal elements of the system, thus it induces dissipative dynamics, while the other couples to diagonal elements, and by itself it generates pure dephasing. By solving the problem analytically exactly, we show that when both baths act concurrently, a strong system-bath coupling gives rise to nonadditive effects in the dynamics. A prominent signature of this nonadditivity is the characteristic {\it slowing down} of population relaxation, driven by the influence of the dephasing bath.
Beyond dynamics, we investigate the thermodynamic behavior of the model. 
Previous studies, using quantum master equations, showed that strong system-bath coupling created bath-cooperativity in this model, allowing heat exchange to the dephasing (diagonally coupled) bath. We find instead that, under the repeated interaction scheme, heat flows exclusively to the dissipative bath (coupled through nondiagonal elements). 
Our results highlight the need for a deeper understanding of the types of open quantum system dynamics and steady-state phenomena that emerge within the repeated interaction framework and the relation of this protocol to other common open quantum system techniques. 
\end{abstract}

\maketitle



\section{Introduction}
\label{sec:intro}

Recent studies have revealed non-trivial cooperative dynamics in systems coupled to two independent baths through non-commuting system operators \cite{Gernot17,AhsanPRL,QUAPI,KamalPRL,Kamal2,Jakub1,Jakub2,Campbell2025}. Consider a two-level system simultaneously interacting with two baths: one couples to nondiagonal elements of the system, thus enabling energy exchange and population relaxation, and the other couples to diagonal system's elements, inducing by itself pure dephasing. When both baths couple strongly to the system, a cooperative phenomenon, often termed frustration, emerges. Specifically, Refs. \cite{Jakub1,Gernot17,AhsanPRL,QUAPI,KamalPRL,Kamal2} identified parameter regimes in which stronger coupling to the dephasing (diagonal) bath slows population relaxation. Conversely, strong coupling to the dissipative (non-diagonal) bath can in a certain parameter regime suppress decoherence \cite{Jakub2}, a result obtained using quantum master equation (QME) approaches. 

The Repeated Interaction (RI) method, also termed the collision model, offers an efficient and transparent framework for modeling open quantum systems (OQSs) \cite{RIrev,Grimmer2016,Ziman2005}. 
In this approach, the environment is represented by a single ancilla, which repeatedly interacts with the system, and is refreshed to its thermal state immediately after a collision with the system. The RI method has been widely applied to study, for example, the thermalization dynamics in OQS \cite{Buzek02,RIMerkli14,RIFilip17,Campbell20,RIZambrini21,SegalRIA,SegalRIC, Barra2023, Parrondo2022}, as well as the operation of OQS as quantum thermal machines \cite{RIMBarra,Strasberg17,Haack1,Haack2,Landi24}. These cited works represent only a small subset of a rapidly growing body of research. 
Using the RI framework to understand baths' cooperativity from first principles and its impact to suppress or accelerate the system relaxation dynamics is one of the motivations for this study, connected to efforts to use thermal baths to engineer and protect quantum states (``dissipation engineering"), see for example Ref. \cite{Murch}. 

Due to its close connection with gate-based quantum computation, the RI framework also provides a promising route for developing quantum algorithms capable of simulating non-unitary dynamics on quantum hardware \cite{Poletti23,Donadi24,Vedral24,Pocrnic25}.
Recently, the RI protocol has been proposed as a framework for preparing quantum thermal states on digital quantum processors \cite{Wiebe2025, SegalRIA, SegalRIC,Buzek02,Koch2025, Abanin2025, Abanin2025_2, Linli2025, Taranto2025, Polla2021,Cubitt2023, Rizzi2024, Maniscalco2023,Donadi24,Campisi2024}. 
Quantum thermal state preparation concerns evolving a quantum system from an arbitrary initial state to a Gibbs thermal state. This process is crucial for applications including the simulation of quantum systems at a given temperature using digital quantum computers \cite{Wiebe2025,Pocrnic25, Chen2023,Goold2021, Donadi24, Maniscalco2023,Linli2025_3,Cubitt2023_2, Rizzi2024,Cubitt2023,Feng2022,Apollaro2024,Arad2025,Alhambra2024, Alhambra2025, Linli2025_2,Shtanko2023} and encoding probability distributions in quantum machine learning \cite{Melko2018, Chen2023}. 
However, even though thermalization appears to be a natural process for a physical quantum system that interacts with a thermal bath, its algorithmic implementation remains challenging, presenting a \textit{QMA complexity} \cite{Chen2023,Shtanko2023}. 
The first description of the quantum thermalization process within the RI framework can be found in the seminal work \cite{Buzek02}, where it was analytically proved that a system qubit can be thermalized by ancilla qubits in the RI scheme if and only if the system-ancilla interaction is energy conserving, that is, the unitary describing the time evolution is a partial swap. In recent studies \cite{Wiebe2025,SegalRIA,SegalRIC,Koch2025}, the energy-conserving condition was relaxed: It was shown that the RI scheme could approximately thermalize a qubit system, and higher dimensional systems, even when the interaction Hamiltonian took a more general energy non-conserving form, as long as collisions were weak but long, corresponding to an effective rotating wave approximation. 
Devising and testing new ideas for thermal state preparation on a quantum hardware based on the RI protocol is a second motivation of this project.


In this study, we use the RI framework to analytically investigate the non-trivial phenomena of suppressing population relaxation with a dephasing bath, and slowing decoherence with a dissipative bath \cite{Jakub1,Jakub2,Gernot17, AhsanPRL, QUAPI}. 
Our goals are threefold. First, while the RI method is a powerful tool for emulating OQS dynamics, the range of dynamical effects that it can capture, compared to QMEs, remains an open question. If the RI framework is to be used to build quantum algorithms for OQS, this question has to be resolved.  
Second, the RI framework allows us to treat strong coupling and non-Markovianity as independent features \cite{RIrev,Campbell20,SegalRIA}, unlike conventional OQS approaches that typically assume both weak coupling and Markovianity simultaneously through the Born-Markov approximation \cite{Petruccione2002}. 
This flexibility makes it possible to disentangle and analyze the distinct roles of strong system–bath coupling and non-Markovian memory effects in both transient dynamics and steady-state behavior.
Third, and specifically for the model studied here, a qubit coupled to two uncorrelated baths via dissipative and dephasing interactions, the RI approach is exactly solvable even at strong coupling. This enables us to gain fundamental insights into the cooperative effects that arise when a system interacts with multiple baths through noncommuting operators  at strong coupling \cite{Jakub1,Jakub2,Cao}, and further provide new ideas for algorithms for thermal state preparation. 


Beyond the questions of relaxation behavior to the steady state and thermalization, we also examine the thermodynamics of the system by analyzing heat exchange and work done at the two system-bath contacts. A key question is whether the RI protocol can show cooperative effects in heat transport, in
line with previous studies of heat exchange at strong coupling, performed using polaron-transformed or reaction coordinate QMEs \cite{Cao,Felix22,HeatMarlon}.

The paper is organized as follows. In Section \ref{sec:model} we present the RI model. In Section \ref{sec:resultsD}, we provide the analytical solution of the dynamics and a full characterization of the steady state, discussing regimes relevant for applications. We also comment on the runtime for thermalizing our model, which is relevant for quantum algorithms development. In Section \ref{sec:thermodynamics}, we present the heat and work exchanged by the system with each of the two baths. We summarize with some open questions in Section \ref{sec:Summ}.

\section{Qubit simultaneously coupled to dissipative and dephasing baths}
\label{sec:model}

\begin{figure}
    \centering
    \includegraphics[width=0.7\linewidth]{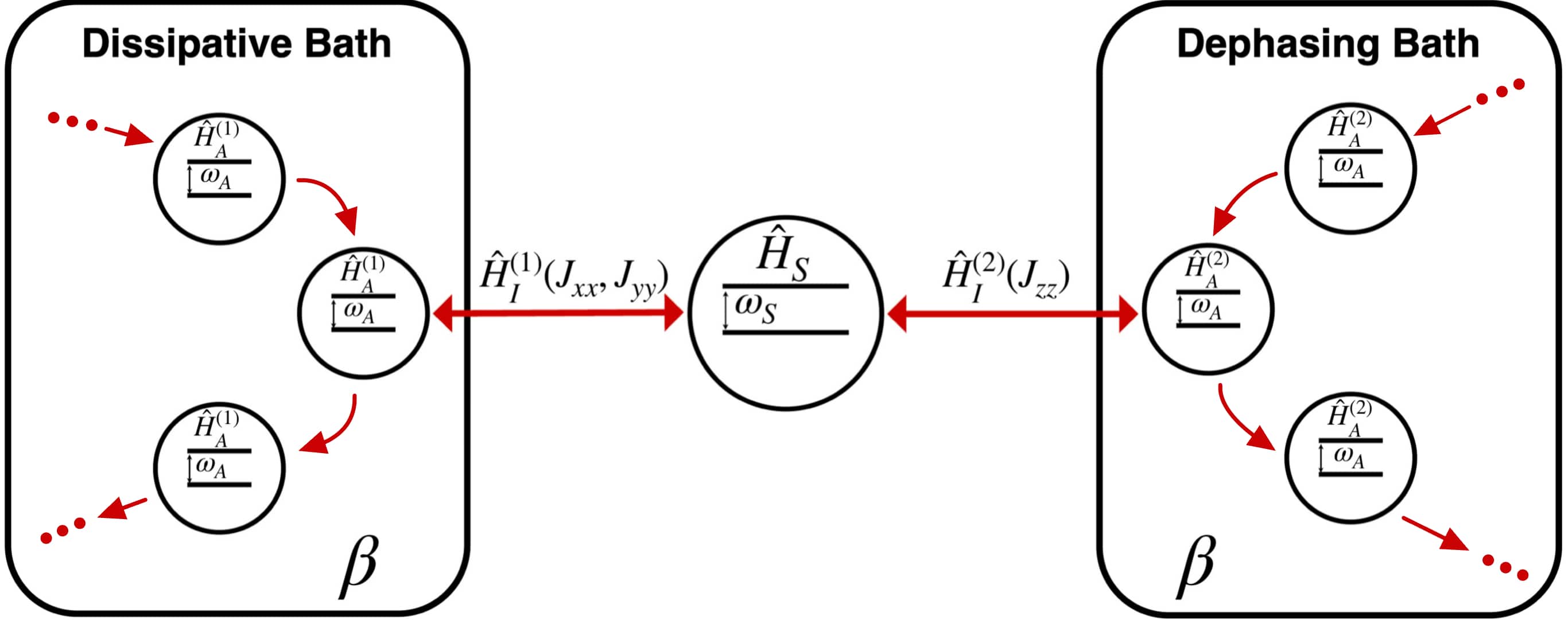}
    \caption{Diagram of the RI model investigated. At each RI step, the system qubit ($S$) interacts simultaneously with both dissipative ($J_{xx}-J_{yy}$) and purely dephasing ($J_{zz}$) baths, at inverse temperature $\beta$, both modeled as a collection of independent and noninteracting qubits, referred to as ancillas ($A$). Each RI step lasts a time interval $\tau$, and the interaction with the dissipative (dephasing) bath is described by the interaction operator $\hat{H}_{I}^{(1)}$ ($\hat{H}_{I}^{(2)}$). After each collision, the ancillas are discarded and replaced by new ones or, equivalently, they are refreshed to their original state.
    }
    \label{fig:system_diagram}
\end{figure}

The model consists a two-level system (qubit) interacting simultaneously with two thermal baths, one couples to nondiagonal elements of the system, the other coupled to diagonal elements. We refer to these baths as ``dissipative" and ``dephasing" baths, respectively. Each bath is modeled as a collection of identical, uncorrelated, and noninteracting ancilla-qubits, prepared in an initial Gibbs thermal state at inverse temperature $\beta$, assumed equal for the two baths. A schematic representation of the model is shown in Fig. \ref{fig:system_diagram}.

The free Hamiltonians of the system (S) and the two baths (A1 and A2) are defined, respectively, as
\begin{equation}
    \begin{split}
        &\hat{H}_{S} = -\frac{\omega_{S}}{2}\hat{\sigma}_{z}^{S},
        \:\:\:\hat{H}_{A}^{(1)} = -\frac{\omega_{A}}{2}\hat{\sigma}_{z}^{A1},
        \:\:\:\hat{H}_{A}^{(2)} = -\frac{\omega_{A}}{2}\hat{\sigma}_{z}^{A2}.
    \end{split}
\end{equation}
%
Here, $\omega_{S}$ and $\omega_{A}$ are the energy splittings, generally different, of the system and the ancillas, respectively; we assume that the two baths have ancillas of equal frequencies. 
The Hamiltonian that describes the interaction between the system and the dissipative bath (A1), $\hat{H}_{I}^{(1)}$, is given by
\begin{equation}
    \hat{H}_{I}^{(1)} = J_{xx}(\hat{\sigma}_{x}^{S}\otimes\hat{\sigma}_{x}^{A1}\otimes\unit^{A2})+J_{yy}(\hat{\sigma}_{y}^{S}\otimes\hat{\sigma}_{y}^{A1}\otimes\unit^{A2}).
\end{equation}
The interaction between the system and the purely dephasing bath (A2), $\hat{H}_{I}^{(2)}$, is given by
\begin{equation}
    \hat{H}_{I}^{(2)} = J_{zz}(\hat{\sigma}_{z}^{S}\otimes\unit^{A1}\otimes\hat{\sigma}_{z}^{A2}).
\end{equation}
The operators $\hat{\sigma}_{x}^{\bullet}$, $\hat{\sigma}_{y}^{\bullet}$, and $\hat{\sigma}_{z}^{\bullet}$ are the $x$, $y$, and $z$ components of the Pauli spin operators, respectively, 
with $\bullet$ standing for S, A1 or A2.
$J_{xx}$, $J_{yy}$, $J_{zz}$ are the coupling strengths along $x$, $y$, and $z$, respectively. The symbol $\unit^{A}$ indicates the identity operator in the Hilbert space of the ancilla $A$. Given these interaction forms, we also refer to the dissipative and purely dephasing baths as the $J_{xx}-J_{yy}$ and the $J_{zz}$ baths, respectively.
At each RI step, the total Hamiltonian, 
$\hat{H}_{tot}$, is 
\begin{equation}
    \hat{H}_{tot} = \hat{H}_{S}\otimes\unit^{A1}\otimes\unit^{A2}
    +\unit^{S}\otimes\hat{H}_{A}^{(1)} \otimes\unit^{A2} +\unit^{S}\otimes\unit^{A1}\otimes\hat{H}_{A}^{(2)}+\hat{H}_{I}^{(1)}+\hat{H}_{I}^{(2)},
\end{equation}
where $\unit^{S}$ represents the identity operator in the Hilbert space of the system $S$.

We assume that the ancillas of each bath are prepared in the same Gibbs thermal state at the same inverse temperature $\beta$,
\begin{equation}
    \rho_{A}^{(1,2)} = 
    \begin{pmatrix}
        p_{A}&c_{A}\\(c_{A})^{*}&1-p_{A}
    \end{pmatrix} = 
    \begin{pmatrix}
        \frac{1}{1+e^{-\beta\omega_{A}}}&0\\0&\frac{e^{-\beta\omega_{A}}}{1+e^{-\beta\omega_{A}}}
    \end{pmatrix},
\end{equation}
Here, $p_{A}$ and $c_{A}$ represent the ground state population and the coherence of the ancillary qubit, respectively. The initial state of the system, $\rho_{S}^{(0)}$, may involve coherences. In the $n^{th}$ RI step, the density matrix that describes the state of the system, $\rho_{S}^{(n)}$, has the general form 
\begin{equation}
    \rho_{S}^{(n)} = 
    \begin{pmatrix}
        p_{S}^{(n)}& c_{S}^{(n)}  \\(c_{S}^{(n)})^{*}&1-p_{S}^{(n)}
    \end{pmatrix}.
\end{equation}
At each time step, the RI dynamics is governed by the map
\begin{equation}
             \rho_{S}^{(n+1)} = \text{Tr}_{A2}\text{Tr}_{A1}\left(\hat{U}(\tau)(\rho_{S}^{(n)}\otimes\rho_{A1}\otimes\rho_{A2})\hat{U}^{\dagger}(\tau)\right),
\label{eq:RI_dynamics}
\end{equation}
where the collision unitary, $\hat{U}(\tau)$, is defined as
\begin{equation}
    \hat{U}(\tau) = e^{-i\hat{H}_{tot}\tau}.
\end{equation}
%
Here, $\tau$ stands for the collision duration, namely the time interval of a single RI step, assumed constant within a process. We set $\hbar=1$ and assume that there is no idle time between two subsequent RI steps. 
Since both the interaction Hamiltonians, $\hat{H}_{I}^{(1)}$ and $\hat{H}_{I}^{(2)}$, and the collision time $\tau$ are fixed, the collision unitary $\hat{U}(\tau)$ is the same at each RI step.

\section{Exact results: Dynamics and steady state}
\label{sec:resultsD}

In this section, we report the exact solution of the model, specifically the analytical expressions describing the population relaxation dynamics, the coherence decay dynamics, and the steady-state values reached by the system.
Our solution is general for any values of the interaction parameters $J_{xx}$, $J_{yy}$ and $J_{zz}$. We then simplify it to the
resonant frequencies and energy-conserving case, namely $\omega_{A}=\omega_{S}$ and $J_{xx}=J_{yy}$. Through numerical and analytical calculations, our objective is to uncover the effect of the $J_{zz}$ bath on the relaxation and decoherence dynamics of the system.

\subsection{Population}

We compute exactly one step of the RI dynamics, Eq. (\ref{eq:RI_dynamics}). We find that the ground state (GS) population of the system follows a recursive relation,
\begin{equation}
    p_{S}^{(n+1)}-p_{S}^{(\infty)} = \eta(p_{S}^{(n)}-p_{S}^{(\infty)}),
\label{eq:population_ansatz}
\end{equation}
where $\eta$ is the relaxation coefficient to the steady state value, indicated by $p_{S}^{(\infty)}$. 
In Appendix \ref{sec:appendix_EOM} we turn Eq. (\ref{eq:population_ansatz}) to a differential equation. 
We extract $\eta$ from the prefactor of $p_{S}^{(n)}$ in the expression obtained from Eq. (\ref{eq:RI_dynamics}), following the procedure in Ref. \cite{SegalRIA},
finding 
\begin{equation}
\begin{split}
        \eta &= 1-\frac{4(J_{xx}-J_{yy})^{2}}{\alpha^2}\sin^{2}\left(\frac{\alpha \tau}{2}\right)(1-p_{A})-\frac{4(J_{xx}-J_{yy})^{2}}{\kappa^2}\sin^{2}\left(\frac{\kappa \tau}{2}\right)p_{A}\\
&-\frac{4(J_{xx}+J_{yy})^{2}}{\nu^2}\sin^{2}\left(\frac{\nu \tau}{2}\right)(1-p_{A})-\frac{4(J_{xx}+J_{yy})^{2}}{\xi^2}\sin^{2}\left(\frac{\xi \tau}{2}\right)p_{A}.
    \end{split}
    \label{eq:eta}
\end{equation}
The coefficient $\eta$ describes the population relaxation dynamics; when $\eta\to 1$, the dynamics is frozen. 
Explicitly, $\eta$ is given in terms of the following microscopic energy parameters,
\begin{equation}
    \begin{split}
        &\xi :=\sqrt{4(J_{xx}+J_{yy})^{2}+(2J_{zz}+\omega_{A}-\omega_{S})^{2}},\\
        &\nu:=\sqrt{4(J_{xx}+J_{yy})^{2}+(2J_{zz}-\omega_{A}+\omega_{S})^{2}},\\
        &\kappa:=\sqrt{4(J_{xx}-J_{yy})^{2}+(-2J_{zz}+\omega_{A}+\omega_{S})^{2}},\\
        &\alpha:=\sqrt{4(J_{xx}-J_{yy})^{2}+(2J_{zz}+\omega_{A}+\omega_{S})^{2}}.
    \end{split}
    \label{eq:energetic_parameters}
\end{equation}
Inspecting the role of $J_{zz}$, we note that it acts as a frequency {\it detuning term}. 
This is to be contrasted to the model considered in Ref. \cite{SegalRIA}, where the system was coupled to a single bath through a full Heisenberg interaction. There, $J_{zz}$ did not affect the population dynamics nor its steady state.

We insert Eq. (\ref{eq:eta}) into Eq. (\ref{eq:population_ansatz}) and solve it for $p_{S}^{(\infty)}$, thus deriving the following steady state value for the system GS population,
\bea
   &&
   p_{S}^{(\infty)} = 
   \nonumber\\
   &&\frac{J_-^{2}(1-p_{A})\left[\kappa^{2}(1-\cos(\alpha\tau))(1-p_{A})+\alpha^{2}(1-\cos(\kappa\tau))p_{A}\right]\xi^{2}\nu^{2}
    }{J_-^{2}\left[\kappa^{2}(1-\cos(\alpha\tau))(1-p_{A})+\alpha^{2}(1-\cos(\kappa\tau))p_{A}\right]\xi^{2}\nu^{2}
    +J_+^{2} \left[\xi^{2}(1-\cos(\nu\tau))(1-p_{A})+\nu^{2}(1-\cos(\xi\tau))p_{A}\right]\alpha^{2}\kappa^{2}}
    \nonumber\\
    &&+
    \frac{
    J_+^{2}p_{A}\left[\xi^{2}(1-\cos(\nu\tau))(1-p_{A})+\nu^{2}(1-\cos(\xi\tau))p_{A}\right]\alpha^{2}\kappa^{2}}{J_-^{2}\left[\kappa^{2}(1-\cos(\alpha\tau))(1-p_{A})+\alpha^{2}(1-\cos(\kappa\tau))p_{A}\right]\xi^{2}\nu^{2}
    +J_+^{2} \left[\xi^{2}(1-\cos(\nu\tau))(1-p_{A})+\nu^{2}(1-\cos(\xi\tau))p_{A}\right]\alpha^{2}\kappa^{2}}.
    \nonumber\\
\label{eq:steadystate}
\eea
For compactness, we adopted here the short notation $J_-=J_{xx}-J_{yy}$ and $J_+=J_{xx}+J_{yy}$. 
%
Equations (\ref{eq:eta})-(\ref{eq:steadystate})
clearly manifest the cooperative nature of the two baths on the dynamics. 
It is also immediately notable that the steady-state solution may deviate from $p_A$. That is, the system does not necessarily reach the state of the ancilla in the steady state limit, depending on the interaction parameters and the RI step, $\tau$. However, when $J_-=0$, we easily confirm that $p_S^{(\infty)}=p_A$, i.e., the system's population reaches the ancilla's.


\subsubsection{Energy conserving and resonance limit, $J_{xx}=J_{yy}$,  $\omega_{A}=\omega_{S}$}
\label{sec:popER}

To gain insight into the impact of the bath $J_{zz}$ on the relaxation dynamics of system, we consider the energy conserving case, that is, we set $J_{xx}=J_{yy}\equiv J_{xy}$ and work under the resonant condition, $\omega_{A}=\omega_{S}\equiv \omega$. In this limit, 
the parameters defined in Eq. (\ref{eq:energetic_parameters}) reduce to $\xi =\nu=2\sqrt{4J_{xy}^{2}+J_{zz}^{2}}$, $\kappa= 2(\omega-J_{zz})$, and  $\alpha=2(\omega+J_{zz})$. We confirm that the steady state solution reduces to thermal, $p_{S}^{(\infty)}=p_{A}$, and we find
that the relaxation coefficient $\eta$ becomes
\begin{equation}
    \begin{split}
        \eta =  1-\frac{4J_{xy}^{2}}{4J_{xy}^{2}+J_{zz}^{2}}\sin^{2}\left(\tau\sqrt{4J_{xy}^{2}+J_{zz}^{2}}\right).
    \end{split}
\label{eq:eta_energyconserving_resonant_limit}
\end{equation}
Equation (\ref{eq:eta_energyconserving_resonant_limit}) highlights that, when $J_{zz}$ increases beyond a certain cutoff value,  $\eta$ progressively tends towards 1 due to the impact of the denominator. Thus, a large $J_{zz}$ slows the relaxation dynamics of populations.
The physical reasoning is that the $J_{zz}$ bath leads to frequency detuning of the system,
thus the Rabi-like dynamics in Eq. (\ref{eq:eta_energyconserving_resonant_limit}) becomes increasingly off-resonant with growing $J_{zz}$.

\begin{figure}[h!]
    \centering
\includegraphics[width=1\linewidth]{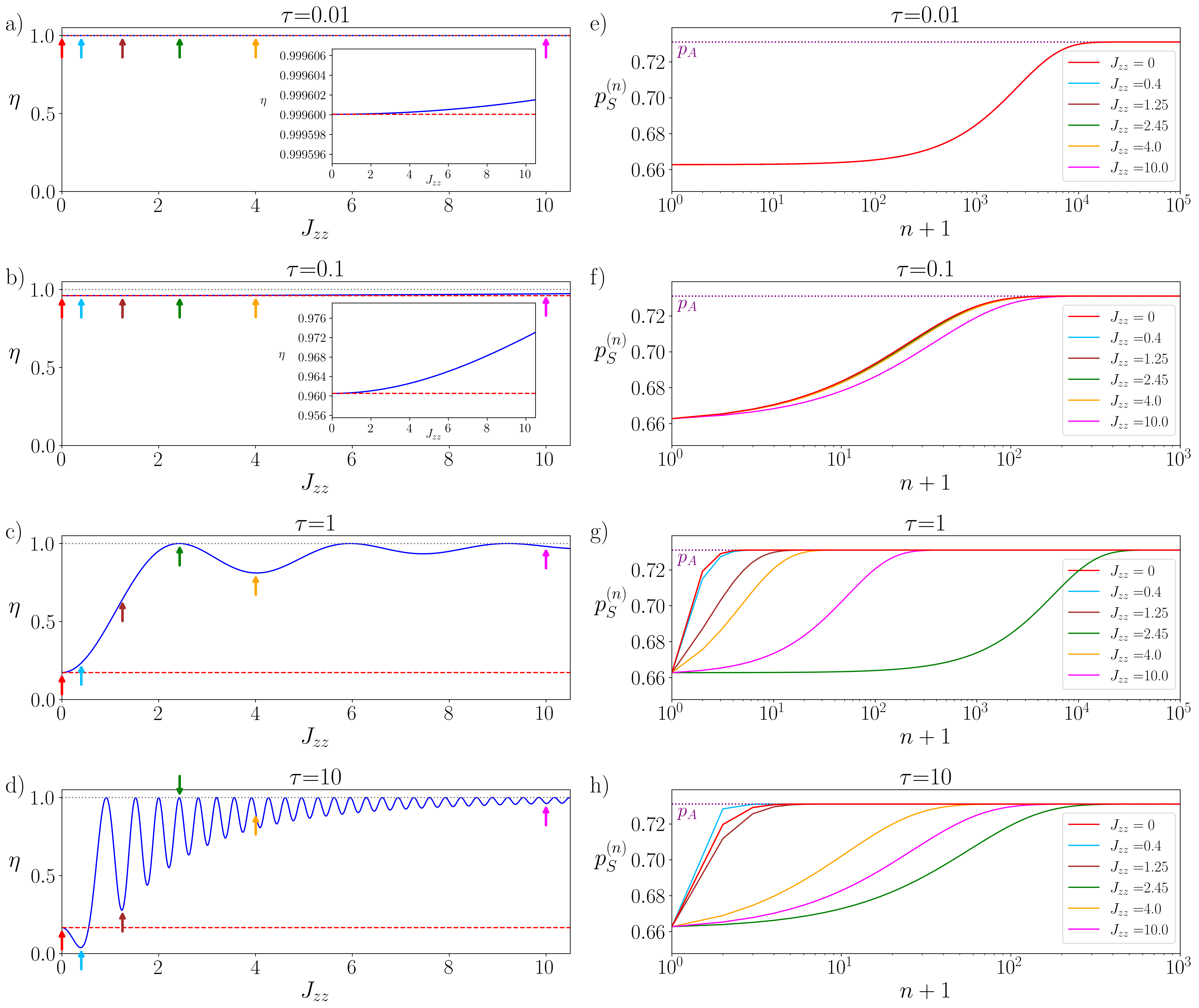}
\caption{Relaxation of the system ground state population as a function of $J_{zz}$. For the choice of interaction parameter,  $J_{zz}$ typically slows down the relaxation dynamics.
(a)-(d) Population relaxation coefficient $\eta$, plotted as a function of $J_{zz}$ for different durations of the RI step, $\tau$. 
The dashed line corresponds to the $J_{zz}=0$ case and the dotted line stands at $\eta=1$ to guide the eye.
The colored arrows point to specific values of $J_{zz}$ that are examined in the corresponding right panels, using the same color coding. 
(e)-(h) Time evolution of the ground state population as a function of the RI step, $n$, for chosen values of $J_{zz}$ and $\tau$. The dotted line represents the ground state population of the ancilla, $p_{A}$. 
We assume energy conserving interactions $J_{xx}=J_{yy}=1$, and work under the resonance condition, $\omega_{A}=\omega_{S}=1$. We use ancillas prepared with $p_{A}=0.73$, corresponding to $\beta=1$.}
\label{fig:eta_vs_Jzz_pop_vs_nstep_differenttau}
\end{figure}

Figure \ref{fig:eta_vs_Jzz_pop_vs_nstep_differenttau} presents this effect: we display $\eta$ and the dynamics of the GS population, $p_{S}^{(n)}$, for different values of $J_{zz}$ and $\tau$. We confine simulations to the resonant and energy conserving limit. 
The behavior of the coefficient $\eta$, Eq. (\ref{eq:eta_energyconserving_resonant_limit}), as $J_{zz}$ is varied while keeping all other parameters fixed is shown in Fig. \ref{fig:eta_vs_Jzz_pop_vs_nstep_differenttau}(a)-(d); the red dashed line corresponds to the $J_{zz}=0$ case.
The corresponding time evolution of the population as a function of the RI step, $n$, is shown in Fig. \ref{fig:eta_vs_Jzz_pop_vs_nstep_differenttau}(e)-(h).
We observe that the common effect of the $J_{zz}$ bath on populations is to {\it slow down the relaxation dynamics}, except in a restricted region of the parameter space typically corresponding to small $J_{zz}$ (see panels (d) and (h)). The suppression of relaxation dynamics becomes more evident as the duration of the RI step, $\tau$, increases. 


\subsubsection{Large $J_{zz}$ limit} 

Generalizing the conclusion from Sec. \ref{sec:popER}, we do not enforce the coupling parameters $J_{xx}$ and $J_{yy}$ to be equal, nor demand resonant frequencies. However, we require that  $J_{zz}\gg J_{xx},J_{yy},\omega_{S},\omega_{A}$. 
In this limit, all the parameters in Eq. (\ref{eq:energetic_parameters}) are approximated by $2J_z$, and we get from Eq. (\ref{eq:eta}) that  
\bea
\eta \approx  1-\frac{J_{+}^2+J_-^{2}}{J_{zz}^{2}}\sin^{2}(J_{zz}\tau). 
\label{eq:Jzzlarge}
\eea
Once again, a strong coupling to the $J_{zz}$ bath generally slows down the relaxation dynamics by increasing $\eta$ towards 1.

\subsubsection{The Stroboscopic-Lindblad limit} 
We consider now the short collision time limit: we assume
$\tau\ll\omega^{-1}$, 
$\tau\ll J_{xx}^{-1}$, but with $J_{xx}^{2}\tau$ a constant, and similarly for $J_{yy}$ and $J_{zz}$. 
In this limit, the RI dynamics is reduced to a Lindblad-type QME \cite{RIrev, Campbell20, Pocrnic25,SegalRIA,Ciccarello2017}. 
We expand the coefficient $\eta$, Eq. (\ref{eq:eta}), in series of $\tau$ around $\tau=0$, stopping at the fourth order, 
\begin{equation}
  \begin{split}
      \eta &= 1-2(J_{xx}^{2}+J_{yy}^{2})\tau^{2}+\frac{1}{12}\left[\alpha^{2}(J_{xx}-J_{yy})^{2}(1-p_{A})+\nu^{2}(J_{xx}+J_{yy})^{2}(1-p_{A})+\right.\\&+\left.
      \kappa^{2}(J_{xx}-J_{yy})^{2}p_{A} +\xi^{2}(J_{xx}+J_{yy})^{2}p_{A}\right]\tau^{4}+O(\tau^{5})\:\:.
  \end{split}
  \label{eq:eta_expanded}
\end{equation}
Notably, $J_{zz}$ appears in the $4^{th}$ order in $\tau$, but not in lower orders. As a result, the impact of the $J_{zz}$ bath on populations dynamics in the short collision time limit is generally imperceptible.

In the resonant energy-conserving limit, i.e., $J_{xx}=J_{yy}\equiv J_{xy}$ and under resonance condition, $\omega_{A}=\omega_{S}\equiv\omega$, Eq. (\ref{eq:eta_expanded}) reduces to
\begin{equation}
    \eta = 1-4J_{xy}^{2}\tau^{2}+\frac{16}{3}J_{xy}^{4}\tau^{4}+\frac{4J_{xy}^{2}}{3}J_{zz}^{2}\tau^{4}+O(\tau^{5})\:\:.
\label{eq:eta_expanded_ec_resonant}
\end{equation}
This expression shows that the $J_{zz}$ bath slows down the relaxation dynamics of the population, making the rate $\eta$ gets closer to $1$.
However, as mentioned above, when $\tau$ is small, this slowing effect is minuscule. In fact, looking at Fig. \ref{fig:eta_vs_Jzz_pop_vs_nstep_differenttau} panels (a) and (e), no appreciable differences is observed in the dynamics between the $J_{zz}=0$ and the $J_{zz}\neq0$ cases since $\tau$ is short.
In panels (c) and (d), we note that, as $\tau$ increases, oscillations in the values of $\eta$ emerge due to the sinusoidal term in Eq. (\ref{eq:eta_energyconserving_resonant_limit}), making the slowing down of populations dynamics more compound, see panels (g) and (h). In particular, when $\tau$ is large
we also find that for a certain range of $J_{zz}$ values, $\eta$ decreases, i.e., populations dynamics {\it accelerates} due to the $J_{zz}$ bath, see panel (d).
%

The fourth order term in Eq. (\ref{eq:eta_expanded_ec_resonant}) exhibits {\it nonadditivity} of the baths, with the contribution $ J_{xy}^{2}J_{zz}^{2}$. 
This effect, also observed in previous studies \cite{Jakub1}, becomes evident only at strong coupling. 
We stress that nonadditivity at strong coupling due the two noncommuting baths is not a prerogative of the short collision time limit or of the resonant energy-conserving limit. In fact, this effect is clear from the general results, Eqs. 
(\ref{eq:eta})-(\ref{eq:energetic_parameters}).

\subsubsection{The $J\tau1$  limit}
\label{sec:Jtau1_population}
We now analyze Eq. (\ref{eq:eta}) in the so-called $J\tau1$ limit \cite{SegalRIA,SegalRIC}. This limit is achieved under three assumptions: (i) $\omega_{A}=\omega_{S}\equiv\omega$ (resonant condition); (ii) $J_{xx},J_{yy},J_{zz}\ll\omega$ (weak-coupling); 
(iii) $J_{nn}\tau$ order of 1, for $n=x,y,z$.
In this limit, the energy parameters defined in Eq. (\ref{eq:energetic_parameters}) simplify to $\alpha,\kappa\approx2\omega, \:\:\xi,\nu \approx 2\sqrt{(J_{xx}+J_{yy})^{2}+J_{zz}^{2}}\:\:$
and $\alpha,\kappa\gg \xi,\nu$.
As a result, $\eta$ reduces to 
\begin{equation}
    \eta\approx1-\frac{ \left(J_{xx}+J_{yy}\right){}^2 \sin ^2\left( \tau \sqrt{\left(J_{xx}+J_{yy}\right){}^2+J_{ zz}^2}\right)}{ \left(J_{xx}+J_{ yy}\right){}^2+J_{ zz}^2}\:\:.
\label{eq:eta_Jtau1}
\end{equation}
Due to the sinusoidal term, analytically evaluating the effect of $J_{zz}$ is challenging.
However, if we consider all couplings to be of similar order $J$ we get 
\bea 
\eta\approx1-\frac{4}{5}\sin^{2}\left(\sqrt{5}J\tau\right). 
\label{eq:etaJtau1a}
\eea 

We compare this result with $\eta_0$, which corresponds to taking $J_{zz}=0$ in Eq. (\ref{eq:eta_Jtau1}),
\bea
  \eta_{0} \approx 1-\sin^{2}(2J\tau).
  \label{eq:eta0}
\eea
In Fig. \ref{fig:eta-eta0_psi-psi_0_Jtau1limit}, we focus on the $J\tau1$ limit, and plot the difference $\eta-\eta_0$ as a function of $J$. We find that whether $J_{zz}$ accelerates ($\eta<\eta_0$) or slows ($\eta>\eta_0$) the relaxation dynamics depends on the coupling energy, with the system alternating between different regimes.

\subsubsection{Summary of observations: population dynamics}

The question we posed was whether dephasing in the form of coupling to the $J_{zz}$ bath slows or accelerates the population relaxation dynamics. 
Slowing down relaxation is beneficial for maintaining quantum states. Accelerating relaxation is important for thermal state preparation applications. 
We largely identified two regimes:

(i) When the interaction terms are order of frequency, the relaxation dynamics can be slowed down by increasing $J_{zz}$, to become the largest energy scale, in accordance with the Zeno dynamics.
This limit is captured by Eq. (\ref{eq:Jzzlarge}) and Fig. \ref{fig:eta_vs_Jzz_pop_vs_nstep_differenttau}.

\begin{figure}[h!]
    \centering
\includegraphics[width=0.6\textwidth]{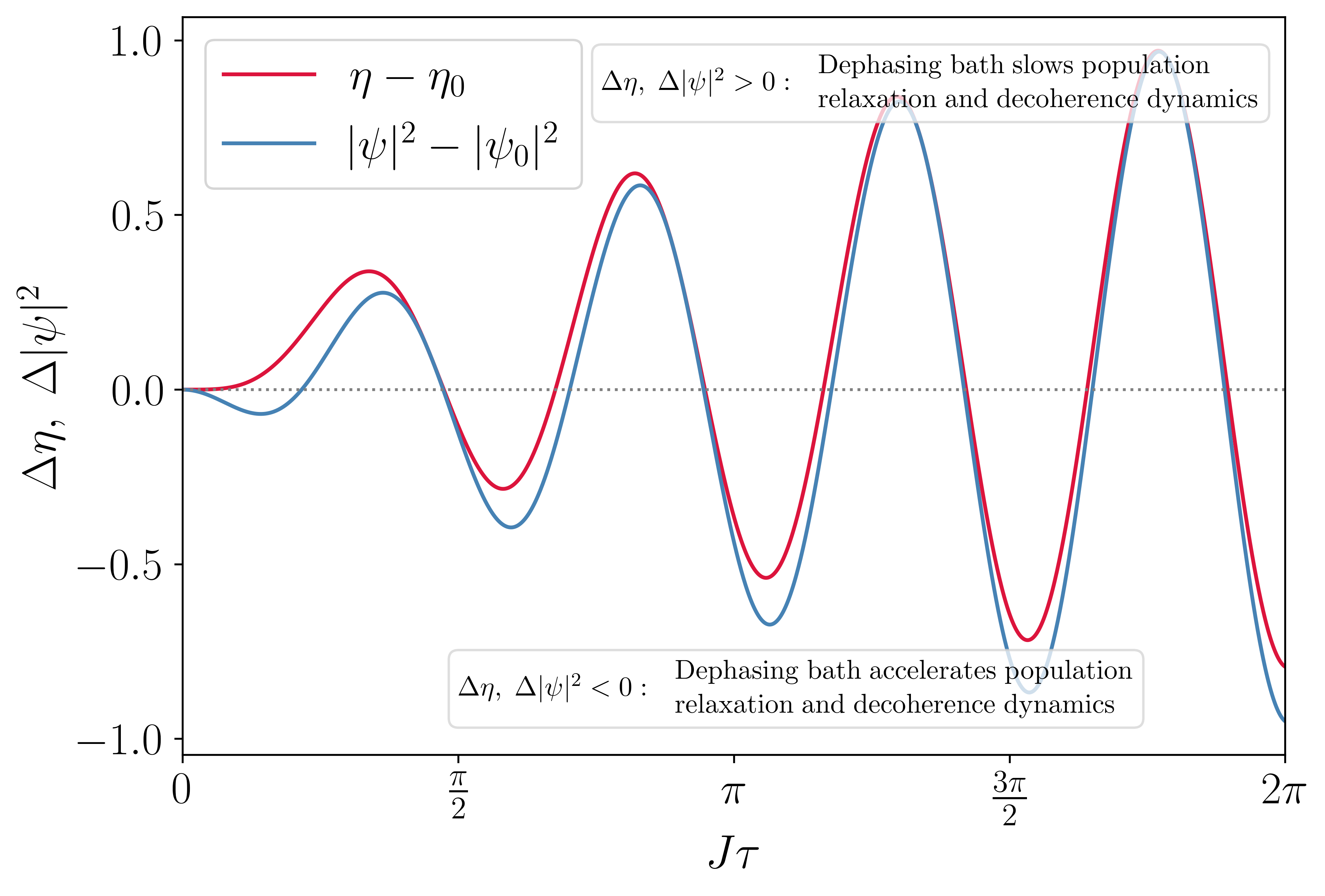}
\caption{Comparison of population decay, $\eta$ and $\eta_{0}$, and coherences decay modulus squared, $|\psi|^{2}$ and $|\psi_{0}|^{2}$, in the $J\tau 1$ limit. 
$\eta$ and $|\psi|^2$ describe dynamics when $J_{zz}\neq0$, while $\eta_0$ and $|\psi_0|^2$
are the $J_{zz}=0$ cases.
$\eta$ ($\eta_0$) is computed using Eq. (\ref{eq:eta_Jtau1}) [Eq. (\ref{eq:eta0})]. 
$|\psi|^{2}$ ($|\psi_{0}|^{2}$)  is computed using Eq. (\ref{eq:psi_square_simpl}) [Eq. (\ref{eq:psi_0})]. 
In simulations we use $J_{xx}=J_{yy} = J_{zz}\equiv J= 10^{-3}$, $\omega_{A}=\omega_{S}=1$, $\tau\in[0,2\pi/J]$, $\beta=1$.}
\label{fig:eta-eta0_psi-psi_0_Jtau1limit}
\end{figure}

(ii) In the limit of weak interactions and long collisions (limit $J\tau1$), the system dynamics can be largely accelerated or slowed down depending on the interaction energy, as predicted by Eqs. (\ref{eq:etaJtau1a})-(\ref{eq:eta0}) and observed in Fig. \ref{fig:eta-eta0_psi-psi_0_Jtau1limit}.
Notably, in this case the coherences closely follow the dynamics of populations, and they are slowed down or accelerated in tandem, as we discuss below.


\subsection{Coherences}

Hand in hand with the relaxation of the population, we should understand the role of $J_{zz}$ in coherence dynamics. 
We now analyze the dynamics of coherences as derived from Eq. (\ref{eq:RI_dynamics}). We use the polar form, $c_{S}^{(n)} = |c_{S}^{(n)}|e^{i\chi}$, $(c_{S}^{(n)})^{*}=|c_{S}^{(n)}|e^{-i\chi}$.
The evolution of coherences in a single RI step satisfies $c_{S}^{(n+1)}=\psi|c_S^{(n)}|$ with
%
\begin{equation}
    \begin{split}
        \psi&= \frac{4 (J_{xx}^{2}-J_{yy}^{2})}{\alpha\kappa\nu\xi} \left[ \left(\kappa\xi\sin\left(\frac{\alpha\tau}{2}\right)\sin\left(\frac{\nu\tau}{2}\right)(1-p_{A})+\alpha\nu\sin\left(\frac{\kappa\tau}{2}\right)\sin\left(\frac{\xi\tau}{2}\right)p_{A}\right) \right]e^{-i\chi}+\\
        &+\left[\left(\cos\left(\frac{\alpha\tau}{2}\right)+\frac{i(2J_{zz}+\omega_{A}+\omega_{S})}{\alpha}\sin\left(\frac{\alpha\tau}{2}\right)\right)\left(\cos\left(\frac{\nu\tau}{2}\right)+\frac{i(2J_{zz}-\omega_{A}+\omega_{S})}{\nu}\sin\left(\frac{\nu\tau}{2}\right)\right)(1-p_{A})\right.\\&+\left.\left(\cos\left(\frac{\xi\tau}{2}\right)-\frac{i(2J_{zz}+\omega_{A}-\omega_{S})}{\xi}\sin\left(\frac{\xi\tau}{2}\right)\right)\left(\cos\left(\frac{\kappa\tau}{2}\right)-\frac{i(2J_{zz}-\omega_{A}-\omega_{S})}{\kappa}\sin\left(\frac{\kappa\tau}{2}\right)\right)p_{A} \right]e^{i\chi}.
    \end{split}
    \label{eq:psi}
\end{equation}
We now ask the question whether the $J_{zz}$ bath accelerates or suppresses the decoherence dynamics.

\begin{figure}[h!]
    \centering
    \includegraphics[width=1\linewidth]{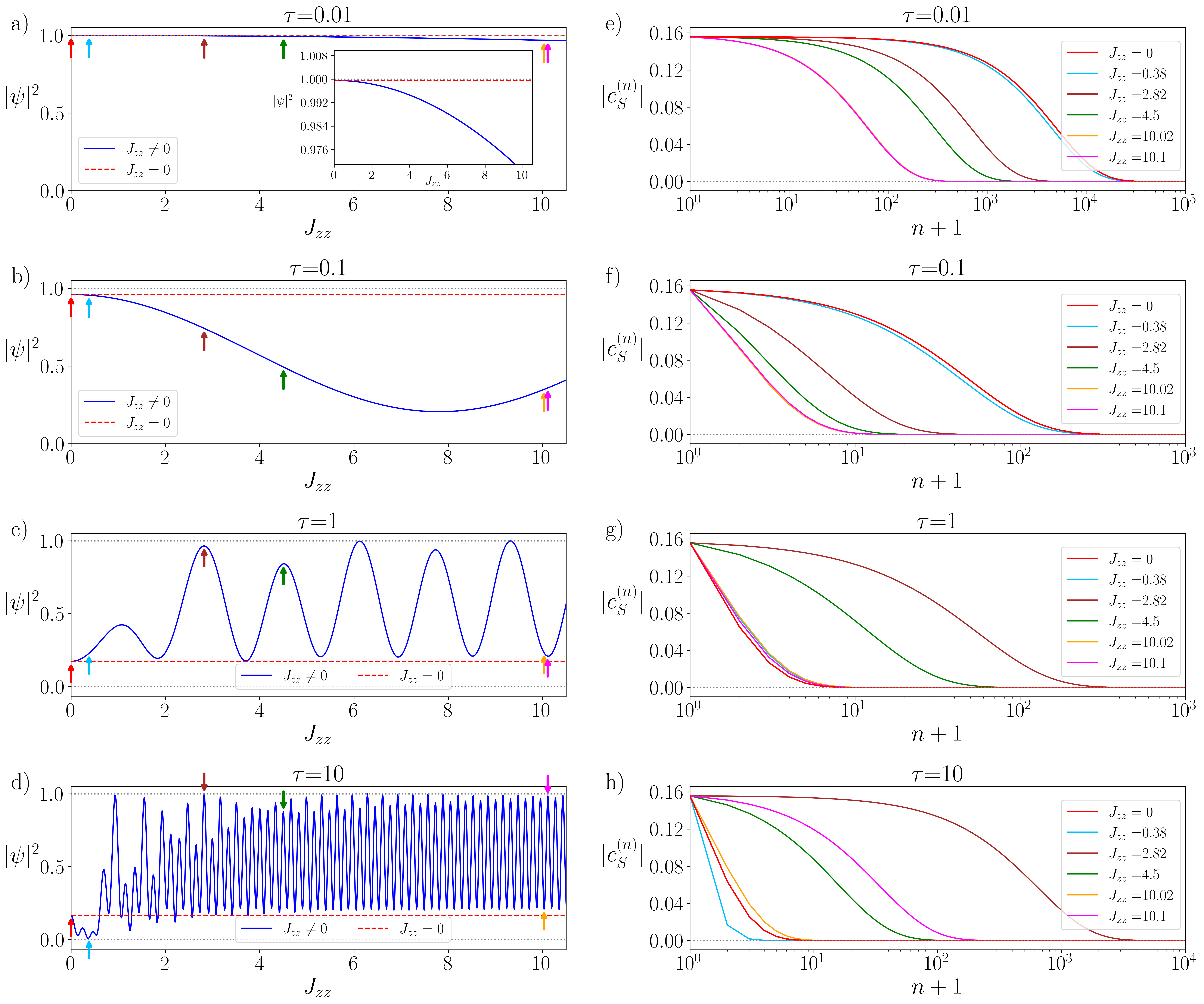}
    \caption{Dynamics of the system's coherences for different values of $J_{zz}$.
    (a)-(d) Amplitude squared of the coherence decay rate, $|\psi|^{2}$, reported as a function of $J_{zz}$ for different durations of the RI step, $\tau$. Colored arrows point to selected couplings strength that are used for the detailed study of dynamics in the right panels, following the same color coding.
    The dashed line corresponds to the $|\psi_0|^2$ case. To guide the eye, the dotted lines mark the values 0 and 1.
(e)-(h) Time evolution of the coherences amplitude, $|c_{S}^{(n)}|$, as a function of the number of RI steps $n$ for different values of $J_{zz}$ and $\tau$. 
We assumed energy conserving interactions, $J_{xx}=J_{yy}=1$, resonant condition, $\omega_{A}=\omega_{S}=1$, and set $p_{A}=0.73$ ($\beta=1)$.}    
\label{fig:psi_mod_vs_Jzz_coh_vs_nstep_differenttau}
\end{figure}

\subsubsection{Energy conserving and resonance limit, $J_{xx}=J_{yy}$, $\omega_{A}=\omega_{S}$}

In the energy conserving limit, $J_{xx}=J_{yy}\equiv J_{xy}$, and under the resonance condition, $\omega_{S}=\omega_{A}\equiv\omega$, 
Eq. (\ref{eq:psi}) reduces to
%
\begin{equation}
\psi= \left[
e^{i(\omega+J_{zz})\tau} \left( \cos\left(\frac{\nu\tau}{2}\right) + \frac{i2J_{zz}}{\nu}\sin \left(\frac{\nu\tau}{2}\right) \right)(1-p_{A})
+e^{i(\omega-J_{zz})\tau} \left(\cos\left(\frac{\nu\tau}{2}\right)-\frac{i2J_{zz}}{\nu}\sin\left(\frac{\nu\tau}{2}\right)\right)p_{A} \right] e^{i\chi}\:\:.
\label{eq:psi_energyconserving_resonant}
\end{equation}
Here, the energy parameters of Eq. (\ref{eq:energetic_parameters}) reduce to $\xi =\nu=2\sqrt{4J_{xy}^{2}+J_{zz}^{2}}$,  $\kappa= 2(\omega-J_{zz})$, and  $\alpha=2(\omega+J_{zz})$.
It is useful to focus on the modulus squared value of the coherences with their decay constant, 
\begin{equation}
\begin{split}
    |\psi|^{2} &= 1-\sin ^2\left(\frac{\nu\tau}{2}\right) +\frac{4(1-2p_{A})^{2}J_{zz}^{2}}{\nu^{2}}\sin ^2\left(\frac{\nu\tau}{2}\right)
    -4\left(1-p_{A}\right) p_{A} \sin^{2} \left(\tau  J_{zz}\right) \left(1-\left(\frac{4 J_{zz}^{2}}{\nu^{2}}+1\right) \sin^{2}\left(\frac{\nu  \tau }{2}\right)\right)\\&-\frac{8 \left(1-p_{A}\right) p_{A} J_{zz} \sin \left(\frac{\nu  \tau }{2}\right) \cos \left(\frac{\nu  \tau }{2}\right) \sin \left(2 \tau  J_{zz}\right)}{\nu }\:\:.
\end{split}
\label{eq:psi_square_simpl}
\end{equation}
When taking $J_{zz}=0$, we get
\begin{equation}
    |\psi_{0}|^{2} = 1-\sin^{2}\left(2J_{xy}\tau\right)\:.
    \label{eq:psi_0}
\end{equation}
Figure \ref{fig:psi_mod_vs_Jzz_coh_vs_nstep_differenttau} presents the decoherence dynamics in this limit
for different values of $J_{zz}$ and $\tau$. The behavior of $|\psi|^{2}$, Eq. (\ref{eq:psi_square_simpl}), as a function of $J_{zz}$ is shown in
panels (a)-(d),
while keeping all other parameters fixed; the red dashed line corresponds to $|\psi_0|^{2}$, Eq. (\ref{eq:psi_0}). 
The time evolution of the decoherence process as a function of the number of RI steps, $n$, 
for different values of $J_{zz}$ and $\tau$, is shown in panels (e)-(h).

We note that for a short RI step ($\tau^{-1}$ larger than all other energies) 
the $J_{zz}$ bath {\it accelerates} the decoherence process, see panels (a),(b),(e),(f), which is the intuitive effect of coupling to a dephasing bath. Conversely, when $\tau$ is outside this regime, decoherence is generally slowed down by the $J_{zz}$ bath, see panels (c),(d),(g),(h).
This trend is counter-intuitive. 
We also note that for long $\tau$, there is a restricted range of $J_{zz}$
that provides a faster decoherence ($|\psi|^2<|\psi_0|^2$), see panels (d) and (h). This region closely corresponds to the same interval on the $J_{zz}$ axis found for populations in Fig. \ref{fig:eta_vs_Jzz_pop_vs_nstep_differenttau}, where the coefficient $\eta$ was reduced by $J_{zz}$, providing a faster relaxation dynamics. Thus, in the long collision time limit, we identify cases where the relaxation of both population and coherences can be accelerated by increasing $J_{zz}$ for the system. 

In Appendix \ref{sec:appendix_modelcomparison}, we present a comparison to a model that includes one bath only, coupled to the system through the Heisenberg interaction form. In that case, while the population dynamics does not change with $J_{zz}$, decoherence is typically slowed by this parameter, except for a small region of parameter space when $\tau\gg\omega^{-1}$.

\begin{figure}
    \centering
    \includegraphics[width=1\linewidth]{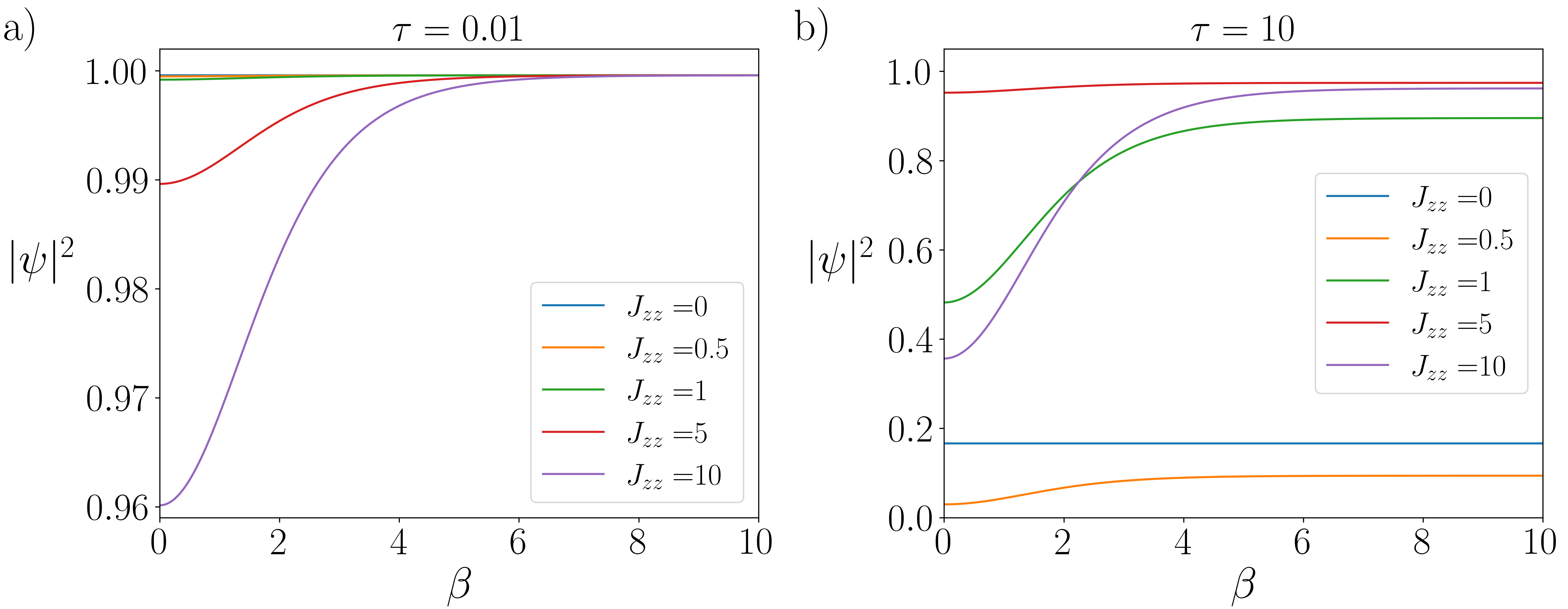}
    \caption{The coherence decay coefficient $|\psi|^{2}$ presented as a function of the inverse temperature of the ancillas, $\beta$, assuming (a) short collisions ($\tau\ll\omega^{-1}=1$) and (b)  long collisions ($\tau\gg\omega^{-1}$). Results were obtained from Eq. (\ref{eq:psi_square_simpl}), setting $J_{xx}=J_{yy}=1,\omega_{A}=\omega_{S}\equiv\omega=1$.}
\label{fig:psi_modsquare_vs_beta}
\end{figure}

In order to clarify the effect of the $J_{zz}$ bath on coherences dynamics, we study now Eq. (\ref{eq:psi_square_simpl}) in different regimes. 

\subsubsection{Large $J_{zz}$ limit}
In the strong decoherence limit, i.e., $J_{zz}\tau\gg1$, $J_{zz}\gg J_{xx}, J_{yy}$, $\omega$,  we have $\nu\approx2J_{zz}$ and thus from Eq. (\ref{eq:psi_square_simpl}) we get
\begin{equation}
    |\psi|^{2} \approx 1-4p_{A}(1-p_{A})\sin^{2}(2J_{zz}\tau)\:\:.
    \label{eq:psi_squar_ec_res_strong_dec}
\end{equation} 
As we indeed see in with Fig. \ref{fig:psi_mod_vs_Jzz_coh_vs_nstep_differenttau}, the decoherence rate oscillates with a frequency that depends on $J_{zz}$ and an amplitude that depends on temperature.
At low temperature, $p_A\to 1$, and the decoherence dynamics gets frozen at large $J_{zz}$ with $|\psi|^2\to 1$. In contrast, at high temperature, $p_A\to 0.5$ and $|\psi|^2$ oscillates with $J_{zz}$ between 0 (fast dynamics) to 1 (frozen dynamics).
Thus, depending on the temperature and the interaction parameters, the decoherence dynamics can be engineered to be either fast or slow.

We present this behavior in Fig. \ref{fig:psi_modsquare_vs_beta}, where we report the value of $|\psi|^{2}$, according to Eq. (\ref{eq:psi_square_simpl}), as a function of the inverse temperature of the ancillas, $\beta$, for (a) short and (b) long collision times in the resonant energy conserving limit. 
The coefficient $|\psi|^{2}$ increases monotonically with $\beta$. When $\tau$ is small, this monotonicity results in frozen dynamics at high $\beta$ for any value of $J_{zz}$, see Fig. \ref{fig:psi_modsquare_vs_beta}(a). In the long collision case, the dynamics can be slowed significantly at low temperature, but only for specific values of $J_{zz}$, and to a lesser extent than at zero temperature, see Fig. \ref{fig:psi_modsquare_vs_beta}(b).
Furthermore, while in the short limit $\tau$, increasing $J_{zz}$ consistently slowed the dynamic, when collisions were made long the role of $J_{zz}$ was more compound, first accelerating the dynamics (compare orange line to blue), then slowing it down (compare green line to blue); see Fig. \ref{fig:psi_modsquare_vs_beta}(b).

\subsubsection{Stroboscopic-Lindblad limit} 

We now study the dynamics in the small collision time limit, arriving at the Lindblad limit, see Appendix \ref{sec:appendix_EOM}. 
We assume for simplicity that $J_{xx}=J_{yy}\equiv J_{xy}$. Expanding Eq. (\ref{eq:psi_square_simpl}) in a series up to the fourth order in $\tau$ around $\tau=0$ we get 
\begin{equation}
    |\psi|^{2} = 1-4J_{xy}^{2}\tau^{2}-16J_{zz}^{2}p_{A}(1-p_{A})\tau^{2}+\frac{64}{3}\tau^{4} \left(1-p_{A}\right)p_{A}J_{zz}^{2}\left(2J_{xy}^{2}+J_{zz}^{2}\right)+\frac{16}{3}J_{xy}^{4}\tau^{4}+\frac{4}{3}J_{xy}^{2}J_{zz}^{2}\tau^{4}+O(\tau^{5})\:. 
    \label{eq:psi_ec_res_shorttime_exp}
\end{equation}
As expected, the effect of the $J_{zz}$ bath on coherences is manifested already at the second order in $\tau$. While the second order term is additive in the baths, the fourth order term reflects a cooperative effect with the product $J_{xy}^{2}J_{zz}^{2}$, a manifestation of strong coupling. 

Focusing on the second order of the expansion, Eq. (\ref{eq:psi_ec_res_shorttime_exp}) indicates that the effect of the $J_{zz}$ bath on decoherence is to {\it speed up} the dynamics, with $|\psi|^2$ becoming smaller as $J_{zz}$ grows. This is intuitive, expected behavior, and consistent with results reported in Fig. \ref{fig:psi_mod_vs_Jzz_coh_vs_nstep_differenttau}, panels (a), (b), (e), (f).
However, as $\tau$ increases, we observe a transition: instead of accelerating the decay of coherences, the $J_{zz}$ bath slows it down, see panels (c), (d), (g), and (h).
This transition is reflected by the next order term in the expansion, switching sign, and increasing $|\psi|^2$.

\subsubsection{The $J\tau1$ limit}

Under the assumptions of the $J\tau1$ limit introduced in Sec. \ref{sec:Jtau1_population}, the energy parameters in Eq. (\ref{eq:energetic_parameters}) simplify to $\alpha$, $\kappa\approx2\omega$, $\xi$, $\nu \approx \sqrt{4(J_{xx}+J_{yy})^{2}+4J_{zz}^{2}}$. From Eq. (\ref{eq:psi_energyconserving_resonant}), we have that
%
%
\begin{equation}
    |\psi|^{2}  =1-\sin^{2}\left(\frac{\nu\tau}{2}\right)\left[1-\frac{J_{zz}^{2}}{4J^{2}+J_{zz}^{2}}(1-2p_{A})^{2}\right].
\end{equation}
Assuming $J_{xx},J_{yy},J_{zz}\sim J$, we get
\bea
|\psi|^{2} \approx 1-\sin^{2}\left(\sqrt{5}J\tau\right)\left[1-\frac{1}{5}(1-2p_{A})^{2}\right],
\eea
which we compare to Eq. (\ref{eq:psi_0}),
$|\psi_{0}|^{2} = 1-\sin^{2}\left(2J\tau\right)$ when nullifying $J_{zz}$.
Comparison of the behavior of coherences in the $J\tau1$ limit with the corresponding population dynamics, Eqs. (\ref{eq:etaJtau1a})-(\ref{eq:eta0}), show that they follow the same trends.
When $p_A\to 1$, the amplitude of the population and the decay of coherence match. As such, accelerated population dynamics due to $J_{zz}\neq 0$ goes hand in hand with faster decoherence rate, and vice versa.
This matching behavior is demonstrated in Fig. \ref{fig:eta-eta0_psi-psi_0_Jtau1limit}.

\subsubsection{Summary of observations: Coherence dynamics}

What is the impact of $J_{zz}$ on the decoherence process?
We identify regimes where population and coherence dynamics go hand in hand, and regimes where they follow opposite trends.

(i) When the interaction energies are order of frequency and $J_{zz}$ is large, we identify regimes of accelerated dynamics under short collision time, and suppressed dynamics, under long collisions.
This rich behavior is demonstrated in Fig. \ref{fig:psi_mod_vs_Jzz_coh_vs_nstep_differenttau}.

(ii) 
In the limit of weak interactions and long collisions (limit $J\tau1$), the system dynamics can be largely accelerated
or slowed down upon scanning the interaction parameters, as observed in Fig. \ref{fig:eta-eta0_psi-psi_0_Jtau1limit}. Importantly, in this regime and at low temperature, population and coherences evolve in tandem, i.e., their relaxation is either accelerated with $J_{zz}$ or slowed.

\subsection{Thermal state preparation: the influence of cooperative effects on runtime $n^{*}$}

The RI scheme can be applied to design algorithms for the preparation of quantum thermal states on a digital quantum processor \cite{Wiebe2025, SegalRIA, SegalRIC,Buzek02,Koch2025, Abanin2025, Abanin2025_2, Linli2025, Taranto2025, Polla2021,Cubitt2023, Rizzi2024, Maniscalco2023,Donadi24}. 
To assess resource needs one needs to estimate the runtime of the algorithm, namely the minimum number of iterations required to bring the system close to the target thermal state, according to some metrics and some threshold $\epsilon$.

In Ref. \cite{SegalRIA}, we showed that using the trace distance as a distance metric, systems prepared in diagonal initial states (e.g. maximally mixed state) required at least $n^{*}$ RI steps to bring the system close to the steady state with an accuracy equal or below a threshold $\epsilon$,
\begin{equation}
    n^{*}\geq \ln\left(\frac{\epsilon}{\left|p_{S}^{(0)}-p_{S}^{(\infty)}\right|}\right)\frac{1}{\ln(|\eta|)}\:\:\:\:.
\label{eq:limit_n*}
\end{equation}
We emphasize that this expression applies when coherences do not exist in the initial condition, nor they are generated during the dynamics.
To handle coherences, numerical simulations should be performed.

\begin{figure}
    \centering
    \includegraphics[width=1\linewidth]{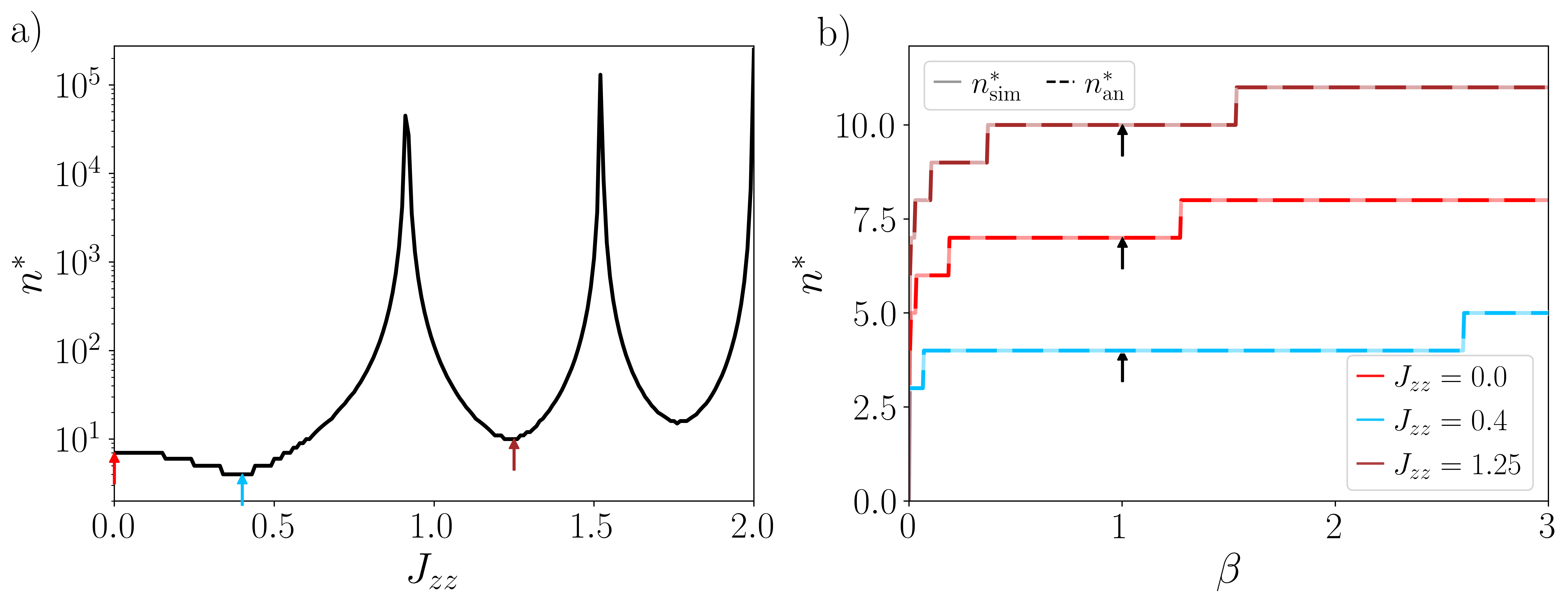}
    \caption{Required number of collisions $n^{*}$ of the RI protocol to achieve thermalization, plotted as function of (a) $J_{zz}$ and (b) $\beta$. In panel (a), we set $\beta=1$. In panel (b), the solid line represents the numerical simulation, $n_{sim}^*$ and the dashed line stands for the theoretical prediction given by Eq. (\ref{eq:limit_n*}), and denoted by $n_{an}*$. We use $J_{xx}=J_{yy}=1$, $\omega_{A}=\omega_{S}=1$, $\epsilon=10^{-6}$, $\tau=10$. We initialize the system in the maximally mixed state.}
    \label{fig:n_star}
\end{figure}

We exemplify this result in
Figure \ref{fig:n_star} working in the resonant and energy conserving limit. We present the runtime $n^{*}$ as a function of (a) $J_{zz}$ and (b) the inverse temperature of the baths, $\beta$, for different values of $J_{zz}$. We observe perfect agreement between numerical simulations (solid) and the prediction (dashed) provided by Eq. (\ref{eq:limit_n*}). 

The runtime $n^{*}$ is consistent with the discussion in Sec. \ref{sec:resultsD} 
and Fig. \ref{fig:eta_vs_Jzz_pop_vs_nstep_differenttau}(d) as
it shows non-monotonicity with respect to $J_{zz}$. In panel (a), peaks in $n^{*}$ are reported; the maximum corresponds to particular combinations of parameters that cause frozen dynamics; in this case, no information is transferred between the system and the two ancillas, resulting in stationary dynamics \cite{SegalRIA}. 

We conclude that the addition of a dephasing bath can allow faster thermalization with respect to the case $J_{zz}=0$ if $J_{zz}$ is tuned properly, here around $J_{zz}=0.4$, consistent with Fig. \ref{fig:eta_vs_Jzz_pop_vs_nstep_differenttau}(d).

\section{Thermodynamics of the model}
\label{sec:thermodynamics}

We demonstrated nontrivial cooperativity of the two baths when considering the dynamics. 
In this section, we study the thermodynamics of our model, providing analytical expressions for heat and work exchanged by the system with each of the two baths. We prove that even though the RI protocol shows cooperative effects in the dynamics, it is unable to capture heat exchange between the system and the $J_{zz}$ bath, even at strong coupling, unlike other OQS methods, e.g., Refs. \cite{Cao,Felix22} .

\subsection{Heat}

The heat exchanged in a single RI step between the system and the $J_{xx}-J_{yy}$ bath is given by
\begin{equation}
    Q^{(n+1)}_{1} = \text{Tr} \left[ \left(\hat{U}(\tau)\hat{H}_{A}^{(1)}\hat{U}(\tau)-\hat{H}_{A}^{(1)}\right)\rho_{S}\otimes\rho_{A1}\otimes\rho_{A2}\right],
\label{eq:heat_dissipative_def}
\end{equation}
where here we define
$\hat{H}_{A}^{(1)}$ over the composite Hilbert space including the system and the second bath.
%
After simplifications, we get
\bea
Q^{(n+1)}_{1} &=& 4\omega_{A}\Bigg[-\frac{(J_{xx}+J_{yy})^{2}}{\nu^{2}}\sin^{2}\left(\frac{\nu\tau}{2}\right)\left(p_{S}^{(n)}-p_{A}\right)(1-p_{A})+\frac{(J_{xx}-J_{yy})^{2}}{\alpha^{2}}\sin^{2}\left(\frac{\alpha\tau}{2}\right)\left(p_{S}^{(n)}-(1-p_{A})\right)(1-p_{A})
\nonumber\\
&+&\frac{(J_{xx}-J_{yy})^{2}}{\kappa^{2}}\sin^{2}\left(\frac{\kappa\tau}{2}\right)\left(p_{S}^{(n)}-(1-p_{A})\right)p_{A}
-\frac{(J_{xx}+J_{yy})^{2}}{\xi^{2}}\sin^{2}\left(\frac{\xi\tau}{2}\right)\left(p_{S}^{(n)}-p_{A}\right)p_{A}\Bigg].
\label{eq:heat_dissipative_expression}
\eea
Similarly, we evaluate the
 heat exchanged in a single RI step between the system and the $J_{zz}$ bath,
\begin{equation}
    Q^{(n+1)}_{2} = \text{Tr} \left[ \left(\hat{U}(\tau)\hat{H}_{A}^{(2)}\hat{U}^{\dagger}(\tau)-\hat{H}_{A}^{(2)}\right)\rho_{S}\otimes\rho_{A1}\otimes\rho_{A2}\right],
\label{eq:heat_exchanged_dephasing bath}
\end{equation}
with $\hat{H}_{A}^{(2)}$ extended to the composite space. 
After simplifications, we get
\begin{equation}
Q^{(n+1)}_{2} = 0,
\end{equation}
indicating that there is no heat exchange between the system and the $J_{zz}$ bath.

We now prove this result directly from Eq. (\ref{eq:heat_exchanged_dephasing bath}). 
To have $Q^{(n+1)}_{2} = 0$ independently of the initial conditions, it must be that
\begin{equation}
     \hat{U}(\tau)\hat{H}_{A}^{(2)}\hat{U}^{\dagger}(\tau)-\hat{H}_{A}^{(2)}=0,
\end{equation}
where $0$ stands for the null operator in the full Hilbert space.
Looking for a non-trivial solution (i.e. $\hat{U}(\tau)\neq 0$), we must show that $\hat{H}_{A}^{(2)}$ and $\hat{U}(\tau)$ commute, namely $\left[\hat{H}_{tot},\hat{H}_{A}^{(2)}\right]\equiv0$.
A direct calculation of this commutator gives
\begin{equation}
\begin{split}
\left[\hat{H}_{tot},\hat{H}_{A}^{(2)}\right]&=\left[\hat{H}_{S}+\hat{H}_{A}^{(1)}+\hat{H}_{A}^{(2)}+\hat{H}_{I},\hat{H}_{A}^{(2)}\right]\\
&=\left[\hat{H}_{I},\hat{H}_{A}^{(2)}\right]\\
&=\left[J_{xx}\hat{\sigma}^{S}_{x}\otimes\hat{\sigma}^{A1}_{x}\otimes\unit^{A2}+J_{yy}\hat{\sigma}^{S}_{y}\otimes\hat{\sigma}^{A1}_{y}\otimes\unit^{A2}
+J_{zz}\hat{\sigma}^{S}_{z}\otimes\unit^{A1}\otimes\hat{\sigma}^{A2}_{z},-\frac{\omega_{A}}{2} \unit^{S}\otimes\unit^{A1}\otimes\hat{\sigma}^{A2}_{z}\right]\\
&=\left[(J_{xx}\hat{\sigma}^{S}_{x}\otimes\hat{\sigma}^{A1}_{x}+J_{yy}\hat{\sigma}^{S}_{y}\otimes\hat{\sigma}^{A1}_{y})\otimes\unit^{A2}+J_{zz}\hat{\sigma}^{S}_{z}\otimes\unit^{A1}\otimes\hat{\sigma}^{A2}_{z},-\frac{\omega_{A}}{2} \unit^{S}\otimes\unit^{A1}\otimes\hat{\sigma}^{A2}_{z}\right]=0,\\
%
\end{split}
\end{equation}
proving our statement. 
This proof is independent of the temperatures of the two baths and this result continues to be valid even when the two baths are maintained at different temperatures. 

\subsection{Work}
The work involved in the interaction between the system and the $J_{xx}-J_{yy}$ bath during a single RI step is defined as
\begin{equation}
    W_{1}^{(n+1)} = \text{Tr}\left[\left(\hat{U}^{\dagger}(\tau)\hat{H}_{I}^{(1)}\hat{U}(\tau)-\hat{H}_{I}^{(1)}\right)\rho_{S}^{(n)}\otimes\rho_{A1}\otimes\rho_{A2}\right],
    \label{eq:work_dissipative_def}
\end{equation}
where $\hat{H}_{I}^{(1)}$
is the interaction Hamiltonian between the system and the $J_{xx}-J_{yy}$ bath \cite{RIMBarra}.
A simplification of Eq. (\ref{eq:work_dissipative_def}) 
provides
\bea
W_{1}^{(n+1)} &=& \frac{4(J_{xx}-J_{yy})^{2}(2J_{zz}-\omega_{A}-\omega_{S})}{\kappa^{2}}\sin^{2}\left(\frac{\kappa\tau}{2}\right)p_{A}\left[p_{S}^{(n)}-(1-p_{A})\right]
\nonumber\\
&+& \frac{4(J_{xx}+J_{yy})^{2}(2J_{zz}+\omega_{A}-\omega_{S})}{\xi^{2}}\sin^{2}\left(\frac{\xi\tau}{2}\right)p_{A}\left(p_{S}^{(n)}-p_{A}\right)
\nonumber\\
&-&\frac{4(J_{xx}-J_{yy})^{2}(2J_{zz}+\omega_{A}+\omega_{S})}{\alpha^{2}}\sin^{2}\left(\frac{\alpha\tau}{2}\right)(1-p_{A})\left[p_{S}^{(n)}-(1-p_{A})\right]    
\nonumber\\
& -&\frac{4(J_{xx}+J_{yy})^{2}(2J_{zz}-\omega_{A}+\omega_{S})}{\nu^{2}}\sin^{2}\left(\frac{\nu\tau}{2}\right)(1-p_{A})\left(p_{S}^{(n)}-p_{A}\right).
\label{eq:work_dissipative_expression}
\eea
Similarly, the work involved in the interaction between the system and the $J_{zz}$ bath during a single RI step is defined as
\begin{equation}
    W_{2}^{(n+1)} = \text{Tr}\left[\left(\hat{U}^{\dagger}(\tau)\hat{H}_{I}^{(2)}\hat{U}(\tau)-\hat{H}_{I}^{(2)}\right)\rho_{S}^{(n)}\otimes\rho_{A1}\otimes\rho_{A2}\right],
    \label{eq:work_dephasing_def}
\end{equation}
where $\hat{H}_{I}^{(2)}$ is the interaction Hamiltonian between the system and the $J_{zz}$ bath.
A simplification of Eq. (\ref{eq:work_dephasing_def}) gives
\begin{equation}
    \begin{split}
        W_{2}^{(n+1)} &= -\frac{4(J_{xx}-J_{yy})^{2}(2J_{zz})}{\kappa^{2}}\sin^{2}\left(\frac{\kappa\tau}{2}\right)p_{A}\left[p_{S}^{(n)}-(1-p_{A})\right]
        \\&- \frac{4(J_{xx}+J_{yy})^{2}(2J_{zz})}{\xi^{2}}\sin^{2}\left(\frac{\xi\tau}{2}\right)p_{A}\left(p_{S}^{(n)}-p_{A}\right)\\&+\frac{4(J_{xx}-J_{yy})^{2}(2J_{zz})}{\alpha^{2}}\sin^{2}\left(\frac{\alpha\tau}{2}\right)(1-p_{A})\left[p_{S}^{(n)}-(1-p_{A})\right]\\
        &+\frac{4(J_{xx}+J_{yy})^{2}(2J_{zz})}{\nu^{2}}\sin^{2}\left(\frac{\nu\tau}{2}\right)(1-p_{A})\left(p_{S}^{(n)}-p_{A}\right). 
    \end{split}
\label{eq:work_dephasing_expression}
\end{equation}
As argued in Ref. \cite{SegalRIA,RIMBarra}, the work involved during each RI step corresponds to the energy supplied to connect the system to the bath at each collision.

\begin{figure}
    \centering
    \includegraphics[width=1\linewidth]{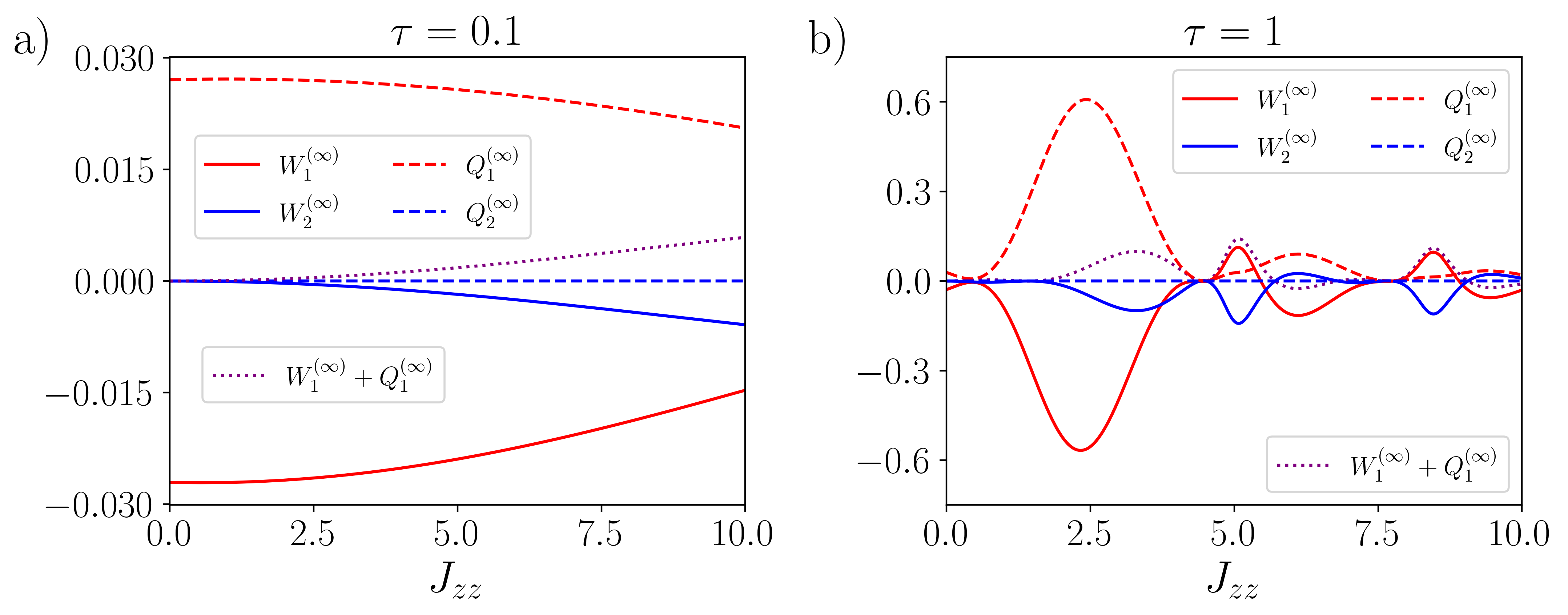}
    \caption{Steady state heat and work exchanged by the system with each bath in a single RI step plotted as a function of $J_{zz}$. Results are presented for two durations of the RI step, (a) $\tau=0.1$ and (b) $\tau=1$. The dotted line represents the total energy exchanged between the system and the $J_{xx}-J_{yy}$ bath. The plots are reproduced directly using the definitions for heat and work reported, Eqs.  (\ref{eq:heat_dissipative_def}), (\ref{eq:heat_exchanged_dephasing bath}), (\ref{eq:work_dissipative_def}), and  (\ref{eq:work_dephasing_def}). We use $J_{xx} = 2$, $J_{yy} = 1$, $\omega_{A} = 2$, $\omega_{S} = 1$, $\beta = 1$. The initial state of the system is $p_{S}^{0} = 0.673$, $c_{S}^{(0)} = 0.45-0.084i$.}
\label{fig:Figure_6}
\end{figure}

Figure \ref{fig:Figure_6} presents the work and heat exchanged by the system with each bath during each RI step in the steady state limit. The results are shown as functions of $J_{zz}$. We consider two different durations of the RI step, and we do not restrict simulations to the resonant energy-conserving case. When $\tau$ increases to $\sim1$, the behavior of heat and work in steady state becomes more compound, and nodes, corresponding to equilibrium states with vanishing heat and work, appear. In addition, variations in work and heat that affect the $J_{xx}-J_{yy}$ bath influence the work incurred at the dephasing contact, and vice versa. This is a clear effect of cooperativity between the two baths.

\section{Summary and open questions}
\label{sec:Summ}

We investigated the repeated interaction scheme as a framework to address the fundamental question of how cooperative effects emerge when an open quantum system interacts simultaneously with two thermal environments through noncommuting system operators. This setting goes beyond the traditional paradigm of additive dissipation and probes the limits of standard assumptions in open quantum system theory. By exactly solving the RI dynamics, we demonstrated that the competition between dissipative and dephasing environments gave rise to nontrivial cooperative phenomena: depending on the parameter regime, the presence of a second bath can either accelerate or suppress relaxation compared to the single-bath scenario. These results revealed the intrinsically nonadditive character of thermalization in the RI framework, highlighting that strong system-bath coupling and noncommutativity could be utilized for quantum state engineering and its preservation.
%
Concretely, we found that while the purely dephasing bath (acting on diagonal elements of the system) typically inhibited relaxation by slowing population dynamics, we identified restricted regions of parameter space, in particular for long interaction times, where it instead enhanced relaxation, allowing resource-efficient thermal state preparation. 
In the stroboscopic-Lindblad limit, cooperative effects disappeared and the dynamics was reduced to the expected additive form of the RI model.
In contrast, in the so-called $J\tau1$ limit, of weak but long collisions, the dynamics alternated between different regimes, of enhanced and suppressed thermalization due to increasing couplings to the bath.


Beyond relaxation dynamics, we examined the thermodynamic behavior of the model. We found that cooperativity did not fully extend to heat exchange: heat flew only through the dissipative-nondiagonally coupled link, while the diagonally coupled bath did not absorb heat. This result contrasts with predictions of QMEs in the strong-coupling regime. The discrepancy points to a fundamental aspect of the RI framework in its current form, namely, the absence of non-Markovian effects. Our results therefore underscore the necessity of incorporating memory effects into RI models at strong coupling in order to capture the full interplay between noncommutativity and strong coupling in open quantum systems.
These insights may also prove valuable for the design of RI-based algorithms to efficiently prepare quantum thermal states on digital quantum computers. 

We now provide a concise summary of our main contributions:

(i) We derived an exact solution for the population and coherence dynamics of a two-level system simultaneously coupled to dissipative and dephasing baths at arbitrary coupling strengths and interaction times. 
The population and coherences of the system manifested cooperative effects at strong system-bath coupling. 

(ii)
We provided an exact analysis of the thermodynamics of the model, proving in general that heat cannot flow in this RI model to the dephasing bath, even at arbitrary strong coupling, contrasting results from QMEs \cite{Cao,HeatMarlon,Felix22}.  

(iii) We demonstrated that the effect of {\it suppression of relaxation by decoherence} was generic to the system once the dephasing coupling energy ($J_{zz}$) was large enough compared to both the dissipative interaction parameters ($J_{xx}$ and $J_{yy}$) and the energy splitting of the system and ancilla. This effect can be interpreted as a Zeno effect \cite{Wineland1990,Chaudry2016,Merkli2025,Zheng2008,KamalPRL,Kamal2}; it was recovered here from first principles. 
Coherences, in contrast, decayed faster due to $J_{zz}$ only at short collision time, but their dynamics was slowed with $J_{zz}$ when collision time was long.

(iv) Working in the so-called $J\tau1$ limit 
\cite{SegalRIA,SegalRIC,Koch2025}, we showed that the dynamics could be suppressed or accelerated by the couplings to the dephasing bath. In fact, it alternated between these regimes when modifying the interaction strengths. 
This trend was exemplified beyond the energy-conserving interaction. The acceleration of the relaxation dynamics when increasing coupling to the environment can be interpreted as an anti-Zeno effect \cite{Pascazio2001,Zheng2008,Gontis1997,Chaudry2016,KamalPRL,Kamal2}. 

(v) We evaluated the thermalization runtime in this two-bath RI protocol, that is, the number of RI iterations required to thermalize the system according to a fixed threshold and adopting the trace distance as a metric. Starting from a completely mixed state, we found that weak coupling to a dephasing bath allowed faster cooling, while strong coupling to that bath increased demands for resources.  

Our results raise a number of open questions about the scope and limitations of the repeated interaction framework. Perhaps the most pressing issue concerns the range of dynamical effects that the RI scheme can faithfully capture. Although its modular structure offers the appealing possibility of disentangling strong coupling from non-Markovianity, it remains unclear what physical phenomena may be lost when memory effects are separated from interaction strength. Understanding these limitations and whether they can be overcome by controlled extensions of the RI model is a key step toward establishing RI as a general framework for open quantum system dynamics.


The thermodynamic aspects of the model opens interesting avenues. We assumed equal bath temperatures, but allowing for temperature differences would drive the system into a nonequilibrium steady state and induce nontrivial energy fluxes. Can strong coupling and cooperative effects be harnessed for thermodynamic tasks such as rectification or energy storage? 

Future research will aim to extend the RI framework by including non-Markovian effects at strong coupling, with the goal of more completely capturing the dynamics of open quantum systems. At the same time, we envision leveraging cooperativity to design RI algorithms for efficient preparation of thermal states on quantum processors.

\begin{acknowledgements}
The work of C. R. E. was supported by the QUARMEN Erasmus Mundus Program and the University of Toronto.
D.S. acknowledges the NSERC Discovery Grant.
The work of C. R. E. and A. P.  was supported by the Department of Physics at the University of Toronto and by the research project: ``Quantum Software Consortium: Exploring Distributed Quantum Solutions for Canada" (QSC). QSC is financed under the National Sciences and Engineering Research Council of Canada (NSERC) Alliance Consortia Quantum Grants \#ALLRP587590-23. 
\end{acknowledgements}
\appendix

\section{Equations of motion in the Stroboscopic-Lindblad limit}

\label{sec:appendix_EOM}

We present here equations of motion for the system population and coherences in the short collision time limit. Consistently with observations reported in Sec. \ref{sec:resultsD}, the equations of motion present nonadditive rates at strong coupling. At weak coupling, the typical additive behavior is restored.

\subsection{Populations}
\label{subsection:apendix_EOM_population}
Based on Eq. (\ref{eq:population_ansatz}),
the equation of motion for the ground state population can be written as
\begin{equation}
   \dot{p}_{S}(t) = -\frac{1-\eta}{\tau}p_{S}(t)+\frac{1-\eta}{\tau}p_{S}^{(\infty)}.
\end{equation}
As can be seen from Eq. (\ref{eq:eta}),
the rate $\frac{1-\eta}{\tau}$ stays constant as $\tau\rightarrow0^{+}$ \cite{SegalRIA}.
Explicitly, Eq. (\ref{eq:eta_expanded_ec_resonant}) is derived under the short collision time limit and assuming resonant frequencies and energy conserving interactions, $\omega_{A}=\omega_{S}$  and $J_{xx}=J_{yy}\equiv J_{xy}$. We then get 
\begin{equation}
  \frac{1-\eta}{\tau} = 4J_{xy}^{2}\tau-\frac{16}{3}J_{xy}^{4}\tau^{3}-\frac{4}{3}J_{xy}^{2}J_{zz}^{2}\tau^{3} + O(\tau^4).
\label{eq:rate_eta_expansion1_resonant_ec}
\end{equation}
%
We now define the rate constants, $\Gamma:=J_{xy}^{2}\tau$ and $\Gamma_{zz}:=J_{zz}^{2}\tau$, assumed constant as $\tau\rightarrow0^{+}$.
Equation (\ref{eq:rate_eta_expansion1_resonant_ec}) then becomes
\bea
\frac{1-\eta}{\tau} &=& 4\Gamma-\frac{16}{3}\Gamma^{2}\tau-\frac{4}{3}\Gamma_{zz}\Gamma\tau.
\eea
This result reflects baths' cooperativity developing at strong coupling. At weak coupling, the rate is given by $\Gamma$, which is dictated by interactions with the dissipative bath only.
At strong couplings, additional terms that depend on interactions of the system with both baths develop, slowing down the process, as reflected by the product $\Gamma\:\Gamma_{zz}$.
%
%

\subsection{Coherences}
%
%
In the short collision time limit, the expansion of $\psi$, Eq. (\ref{eq:psi}) to the third order in $\tau$ around $\tau=0$, provides
\begin{equation}
    \begin{split}
            \psi&= e^{i\chi}\left[1+i(2J_{zz}(1-2p_{A})+\omega_{S})\tau-\left(J_{xx}^{2}+J_{yy}^{2}+2J_{zz}(J_{zz}+(1-2p_{A})\omega_{S})+\frac{\omega_{S}^{2}}{2}\right)\tau^{2}\right.\\&
        \left.-i\frac{1}{6}\left[8J_{zz}(J_{xx}^{2}+J_{yy}^{2}+J_{zz}^{2})(1-2p_{A})+4J_{xx}J_{yy}\omega_{A}+4(J_{xx}^{2}+J_{yy}^{2}+3J_{zz}^{2})\omega_{S}+6J_{zz}(1-2p_{A})\omega_{S}^{2}+\omega_{S}^{3}\right]\tau^{3} \right]\\&
        +e^{-i\chi}(J_{xx}^{2}-J_{yy}^{2})\tau^{2}+O(\tau^{4}).
    \end{split}
\label{eq:psi_short_time}
\end{equation}
In the resonant limit and when assuming energy-conserving interactions we get 
\begin{equation}
    \begin{split}
            \psi&= e^{i\chi}\left[1+i(2J_{zz}(1-2p_{A})+\omega)\tau-\left(2J_{xy}^{2}+2J_{zz}(J_{zz}+(1-2p_{A})\omega)+\frac{\omega^{2}}{2}\right)\tau^{2}\right.\\&
        \left.-i\left(\frac{8}{3}J_{xy}^{2}J_{zz}(1-2p_{A})+\frac{4}{3}J_{zz}^{3}(1-2p_{A})+2\omega(J_{xy}^{2}+J_{zz}^{2})+J_{zz}(1-2p_{A})\omega^{2}+\frac{\omega^{3}}{6}\right)\tau^{3}\right],
    \end{split}
\label{eq:psi_short_time_res_ec}
\end{equation}
where in Eq. (\ref{eq:psi_short_time_res_ec})  higher-order terms in $\tau$ are neglected.
Using $c_{S}^{(n+1)}=\psi|c_{S}^{(n)}|$ and $c_{S}^{(n)}=|c_{S}^{(n)}|e^{i\chi}$, we write
\begin{equation}
\begin{split}
        \frac{c_{S}^{(n+1)}-c_{S}^{(n)}}{\tau} &=\frac{\psi e^{-i\chi}-1}{\tau}c_{S}^{(n)}\\& = \left[i(2J_{zz}(1-2p_{A})+\omega)-\left(2J_{xy}^{2}+2J_{zz}(J_{zz}+(1-2p_{A})\omega)+\frac{\omega^{2}}{2}\right)\tau\right.\\&
        \left.-i\left(\frac{8}{3}J_{xy}^{2}J_{zz}(1-2p_{A})+\frac{4}{3}J_{zz}^{3}(1-2p_{A})+2\omega(J_{xy}^{2}+J_{zz}^{2})+J_{zz}(1-2p_{A})\omega^{2}+\frac{\omega^{3}}{6}\right)\tau^{2}\right] c_{S}^{(n)}.
\end{split}
\label{eq:coh_ec_res_EOM}
\end{equation}
As a side note, the coefficient $\psi$ also depends on $\tau$ through the phase dependence, $\chi=\chi(n)$. However, for notational simplicity, we do not write this dependence explicitly.
Defining once again
$\Gamma:=J_{xy}^{2}\tau$, $\Gamma_{zz}:=J_{zz}^{2}\tau$, and $\Omega:=\frac{\omega^{2}}{2}\tau$, such that these coefficients remain constant as $\tau\rightarrow0^{+}$, 
Eq. (\ref{eq:coh_ec_res_EOM}) becomes
%
\begin{equation}
\begin{split}
    \frac{c_{S}^{(n+1)}-c_{S}^{(n)}}{\tau} &= \Bigg[ i(2J_{zz}(1-2p_{A})+\omega)-(2\Gamma+\Delta+\Omega) \\&-i\left(\frac{8}{3}\Gamma J_{zz}(1-2p_{A})+\frac{4}{3}\Gamma_{zz}(1-2p_{A})J_{zz}+2\omega(\Gamma+\Gamma_{zz})+2J_{zz}(1-2p_{A})\Omega+\frac{\omega}{3}\Omega\right)\tau\Bigg]c_{S}^{(n)},
\end{split}
\end{equation}
with $\Delta:=2J_{zz}(J_{zz}+\omega(1-2p_{A}))\tau$ a constant.
This equation reflects nonadditivity with the $\sim \Gamma J_{zz}$ term affecting the phase.
In the weak coupling limit, when $\tau\rightarrow0^{+}$ we get
\begin{equation}
    \dot{c}_{S}(t) = \left[ i(2J_{zz}(1-2p_{A})+\omega)-(2\Gamma+\Delta+\Omega)\right]c_{S}(t).
    \label{eq:coh_ec_res_EOM_diff}
\end{equation}
%
%
We note that the $J_{zz}$ bath modifies the phase and increases the decoherence rate, adding to the impact of the dissipative bath, $\Gamma$.

\section{Comparison between the two bath model and the Heisenberg single bath model}
\label{sec:appendix_modelcomparison}

We compare here our results with a system coupled to two baths, dissipative and dephasing, to the single bath model with a full Heisenberg interaction, $J_{xx}\neq0$, $J_{yy}\neq 0$, and $J_{zz}\neq0$, as analyzed in Ref. \cite{SegalRIA}. 
Importantly, in the Heisenberg model $J_{zz}$ has no impact on populations, contrasting our results in Sec. \ref{sec:resultsD}.
However, the decoherence dynamics differs between the two models. In the two-bath model, decoherence is governed by the coefficient $\psi$, Eq. (\ref{eq:psi}); in the  Heisenberg interaction model, the decay rate of coherences is given by \cite{SegalRIA},
\begin{equation}
\begin{split}
\tilde{\psi}_{0}=&
\frac{4(J_{xx}^{2}-J_{yy}^{2})}{\theta\phi}e^{-i(\chi+2J_{zz}\tau)}\sin\left(\frac{\theta\tau}{2}\right)\sin\left(\frac{\phi\tau}{2}\right) 
\nonumber\\
+& e^{i(\chi+2J_{zz}\tau)}\left(\cos\left(\frac{\theta\tau}{2}\right)-i\frac{\left(\omega_{A}-\omega_{S}\right)}{\theta}\sin\left(\frac{\theta\tau}{2}\right)\right)\left(\cos\left(\frac{\phi\tau}{2}\right)+i\frac{\left(\omega_{A}+\omega_{S}\right)}{\phi}\sin\left(\frac{\phi\tau}{2}\right)\right) 
\nonumber\\
+& 2ip_{A}\Biggl[\frac{4(J_{xx}^{2}-J_{yy}^{2})}{\theta\phi}e^{-i\chi}\sin\left(\frac{\theta\tau}{2}\right) \sin\left(\frac{\phi\tau}{2}\right)  - 
    e^{i\chi}\left(\cos\left(\frac{\theta\tau}{2}\right)- \frac{i(\omega_{A}-\omega_{S})}{\theta}\sin\left(\frac{\theta\tau}{2}\right)\right)\cdot 
\nonumber\\\cdot&\left(\cos\left(\frac{\phi\tau}{2}\right) + \frac{i(\omega_{A}+\omega_{S})}{\phi}\sin\left(\frac{\phi\tau}{2}\right)\right) \Biggr]\sin(2J_{zz}\tau), 
\label{eq:psi_tilde}
\end{split}
\end{equation}
with $\theta=\sqrt{(J_{xx}+J_{yy})^{2}+(\omega_{A}-\omega_{S})^{2}}$ and $\phi=\sqrt{(J_{xx}-J_{yy})^{2}+(\omega_{A}+\omega_{S})^{2}}$. 
In general,  $|\psi|\neq |\tilde{\psi}_{0}|$.

\begin{figure}[h!]
            \centering
\includegraphics[width=1.0\linewidth]{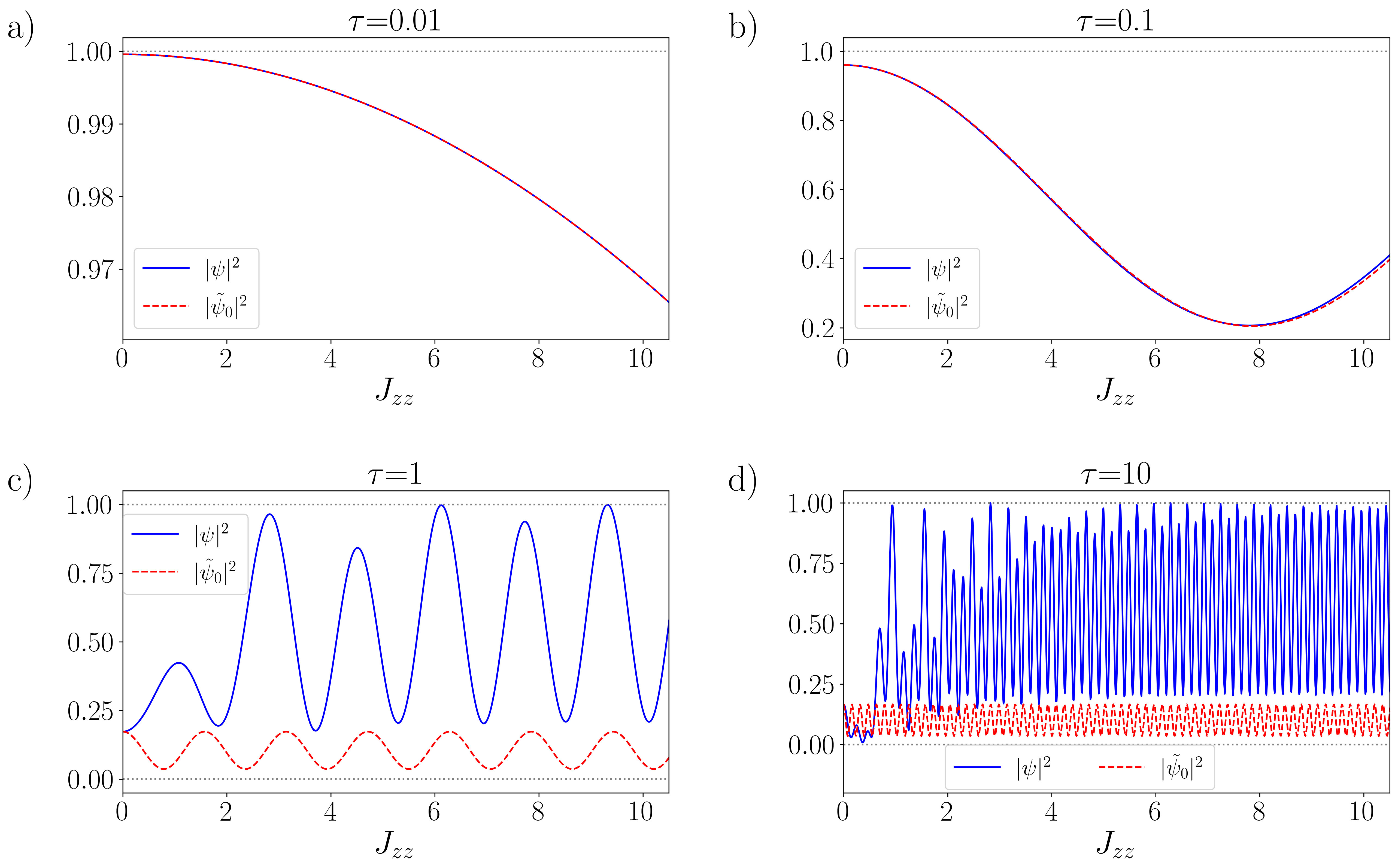}
\caption{Comparison of coherence decay rate amplitude as a function of $J_{zz}$ in the two-bath (dissipative-dephasing) model and in the single bath Heisenberg model, for different values of collision time $\tau$. Parameters are  $J_{xx}=J_{yy}=1$, $\omega_{A}=\omega_{S}=1$, and $\beta=1$. Dotted lines guide the eye to the 0 and 1 values.
}
\label{fig:comparison_psi_psi0tilde}
\end{figure}

Figure  \ref{fig:comparison_psi_psi0tilde} provides a comparison between $|\psi|^{2}$ and $|\tilde{\psi}_{0}|^{2}$ as a function of $J_{zz}$ for different collision times $\tau$. When $\tau$ is small, we observe no significant differences between the models, as expected from considerations made on the Lindblad limit in Sec. \ref{sec:resultsD}.
As $\tau$ increases, 
this difference is no longer negligible and we note that typically $|\psi|^{2}\geq|\tilde{\psi}_{0}|^{2}$ 
for any $J_{zz}$, corresponding to a slower decoherence in the two-bath model. 
When $\tau$ is long,  decoherence in the two-bath model can be faster than under the Heisenberg model only if $J_{zz}$ is tuned properly; this acceleration, however, is modest, see the small $J_{zz}$ range in panel (d).

\end{document}